\documentclass[aps,prd,11pt,preprint,nofootinbib,showkeys]{revtex4-1}
\usepackage{amsmath,amssymb,amsthm,mathrsfs,bbm}
\usepackage{latexsym,amscd,amsbsy,amsfonts,dsfont}
\usepackage{graphicx}
\usepackage[utf8]{inputenc}
\usepackage{subfigure}  
\usepackage{float}
\usepackage{slashed}
\usepackage{cancel}
\usepackage{multirow}
\usepackage{soul}
\usepackage{physics}
\usepackage{comment}
\usepackage[normalem]{ulem}
\usepackage[usenames,dvipsnames]{xcolor}
\usepackage[
      colorlinks=true,
      linktocpage=true,
      breaklinks=true,
      linkcolor=Aquamarine,
      urlcolor=Purple,
      filecolor=black,
      citecolor=Purple,
      menucolor=black,
      pdfstartview=FitV,
      bookmarksopen=true
      ]{hyperref}

\makeindex

\begin{document}

\title{Beyond Buchdahl's limit: bilayered stars and thin-shell configurations}

\author{Julio Arrechea}
\email{julio.arrechea@sissa.it}
\affiliation{SISSA, Via Bonomea 265, 34136 Trieste, Italy}
\affiliation{INFN Sezione di Trieste,
     via Valerio 2, 34127 Trieste, Italy}
\affiliation{IFPU - Institute for Fundamental Physics of the Universe, Via Beirut 2, 34014 Trieste, Italy}
\author{Carlos Barcel\'o}
\email{carlos@iaa.es}
\affiliation{Instituto de Astrof\'{\i}sica de Andaluc\'{\i}a (IAA-CSIC), Glorieta de la Astronom\'{\i}a, 18008 Granada, Spain}
\author{Gerardo Garc\'{\i}a-Moreno}
\email{ggarcia@iaa.es}
\affiliation{Instituto de Astrof\'{\i}sica de Andaluc\'{\i}a (IAA-CSIC), Glorieta de la Astronom\'{\i}a, 18008 Granada, Spain}
\author{Jos\'e Polo-G\'omez}
\email{jpologomez@uwaterloo.ca}
\affiliation{Department of Applied Mathematics, University of Waterloo, Waterloo, Ontario, N2L 3G1, Canada}
\affiliation{Institute for Quantum Computing, University of Waterloo, Waterloo, Ontario, N2L 3G1, Canada}
\affiliation{Perimeter Institute for Theoretical Physics, Waterloo, Ontario, N2L 2Y5, Canada}

\begin{abstract}
    One of the theoretical motivations behind the belief that black holes as described by general relativity exist in nature is that it is hard to find matter configurations that mimic their properties, especially their compactness. One of the classic results that goes in this direction is the so-called Buchdahl limit: a bound for the maximum compactness that spherically symmetric isotropic fluid spheres in hydrostatic equilibrium can possibly achieve with an outward-decreasing energy density. However, physically realistic situations could violate both isotropy and the monotonicity of the density profile. Notably, Bondi already showed that if the density profile is allowed to be arbitrary (but remains non-negative), a less restrictive compactness bound emerges. Furthermore, if negative energy densities are permitted, configurations can approach the black hole compactness limit arbitrarily closely. In this work we introduce a set of simple bilayered and thin-shell toy models designed to illustrate the effect of relaxing separately the assumptions of Buchdahl's theorem. Within these models we highlight the existence of two special examples that we have called AdS stars and Einstein Static stars. We also discuss how these toy models may represent some of the main features of realistic systems, and how they could be extended to find more refined models.
\end{abstract}

\keywords{}

\maketitle
 
\tableofcontents

\section{Introduction}
\label{Sec:Introduction}

Verifying whether the dark and compact objects that we observe in the universe strictly correspond to general relativistic black holes is still an open question that is attracting a large amount of interest and effort~\cite{Werneretal1998,Abramowicz2002,Carballo-Rubio2022}. 
In an astrophysical context, any object that displays a sufficiently large gravitational mass in a sufficiently small region is believed to be a black hole. 
Probing and analyzing astrophysical black hole systems could be our best chance to find a robust path beyond standard general relativity.

On theoretical grounds, the first result pointing out that black holes might be inevitable can be traced back to Chandrasekhar's mass limit\footnote{This limit was first found by Edmund Stoner~\cite{Nauenberg2008}.} for white dwarf stars~\cite{Chandrasekhar1931a,Chandrasekhar1931b}. As a dwarf star increases its mass, the electrons become more relativistic, and a mass limit appears at which the pressure provided by Pauli's exclusion principle for fermions is unable to counteract gravity. An equivalent limit appears for neutron stars, now with the exclusion applied to neutrons~\cite{OppenheimerVolkoff1939}. In both cases, adding some more mass to the limiting configuration would cause the star to collapse into a black hole. It is interesting to recall that these limits are found even without invoking pure general relativistic effects. 

Making use of the full Einstein equations one can find, in addition to the previous limits, some bounds to the compactness of those stars~\cite{NilssonUggla2000}. We shall define the compactness $C(R)$ of a spherically symmetric configuration to be the quotient between twice the total mass of the object $2M$ and its radius $R$: $C(R)=2M/R$. For example, typical neutron stars have compactness of around $C(R) \sim 0.4$. This is to be contrasted with the maximum compactness attainable in general relativity: the limit $C(R)=1$ representing  a black hole, i.e. a completely collapsed star. 

Now, armed with the full Einstein equations one can analyze whether there could be, at least in principle, stellar objects more compact than neutron stars. Here comes the surprise. Under mild conditions there exists a limit to the maximum compactness attainable by a static, spherically symmetric sphere of perfect fluid matching the Schwarzschild metric at its surface. This so-called Buchdahl limit~\cite{Buchdahl1959,Bondi1964,Wald1984} is $C(R)<8/9$, and its origins can be traced back to Schwarzschild's work on stellar interiors~\cite{Schwarzschild1916b}. Buchdahl proved that $C(R)<8/9$ as a theorem, making no strong hypothesis regarding the equation of state of matter beyond being barotropic. Its two central assumptions are just the isotropy of the fluid, namely that pressures in the tangential and radial directions are equal, and that the energy density is a positive, (outwards) monotonically decreasing function of the radius. The same bound seems to hold for the Einstein-Vlasov system~\cite{Andreasson2006}.

Even when the density profile is not required to be a monotonically decreasing function of the radius, it is still possible to establish upper bounds on compactness, provided the density remains non-negative. Bondi~\cite{Bondi1964} found for this situation a general upper bound of \mbox{$C(R) < 12 \sqrt{2} - 16 \approx 0.971$}, which is less restrictive than the Buchdahl bound \mbox{$C(R)<8/9 \approx 0.889$}, but still well below the black hole compactness $C(R)=1$ 
Bondi then proposed a model to illustrate how this limiting compactness could be attained. He vaguely describes this model as a core with no energy density joined to an external Schwarzschild geometry through a thin shell. However, he did not describe the characteristics of this thin shell. For instance, generically a thin shell obeying Israel's junction conditions~\cite{Israel1966} contains tangential pressures which would introduce anisotropic pressures, against the very hypotheses of the set-up. 

Bondi~\cite{Bondi1964} also demonstrated that if negative energy densities are permitted, the compactness can approach the black hole limit as closely as desired. Again, the specific model used to illustrate this result is based on a thin-shell construction that is not fully characterized. Less restrictive results but under more general conditions have also been reported in the literature. For instance, a bound of the form $2m(r)/r < 1$ was found in~\cite{Baumgarte1993} assuming positive energy density and isotropy (see~\cite{Mars1996} for an interesting extension to the anisotropic case, where only regularity at $r=0$ and the condition $\rho + p_r + 2 p_t \geq 0$ are assumed). In~\cite{Karageorgis2007}, Karageorgis and Stalker present a variety of sharp bounds for the compactness of spherically symmetric configurations. On top of adding rigor to Bondi's proof for the isotropic case, they also found other bounds by relaxing isotropy but imposing energy-like conditions.

In this paper, we further explore the consequences and the physical viability of relaxing each individual assumption behind Buchdahl's theorem within a spherically symmetric setup. In particular, we are interested in finding simple and realistic physical models that allow us to understand how relaxing these assumptions allows to bypass the limit. We restrict ourselves to static and spherically symmetric configurations, since potential generalizations to stationary and axisymmetric scenarios are far from straightforward.

First, we will explore the role of the monotonicity and positivity of the energy density by introducing a stellar model composed of two thick layers of constant energy density: an internal core and an external crust, with the condition that the density is always larger for the crust than for the core. For this model, we analyze two very different situations: one in which we constrain the core density to be positive, and another in which we allow it to take negative values. We show that, in the first situation, we find that we are able to approach numerically Bondi's bound $C(R) <  12 \sqrt{2} - 16$ as much as desired. We will relate our models to Bondi's, providing a clearer interpretation of his approach. We then relax the positivity condition by allowing for negative densities in the core. In this case, we show that it is possible to build solutions as close to the black hole limit as desired, at least as long as we do not impose any energy condition on the fluid, in agreement with Bondi's results again.

Within these bilayered models we find one specially simple and interesting: a stellar structure that we have decided to call {\em anti-de Sitter star}. It is similar to a gravastar~\cite{MazurMottola2023} but with an anti-de Sitter (AdS) core instead of a de Sitter core. These stars have also some similarities with AdS black shells~\cite{Danielssonetal2017}, the central difference being precisely their compactness. While AdS black shells have compactness around the Buchdahl limit, AdS stars have compactness close to the black hole limit.  
As we will discuss, it is interesting to highlight that these AdS stars appear as idealizations of the semiclassical stellar configurations found in~\cite{Arrecheaetal2021,Arrecheaetal2023}.

Secondly, as a separate situation, we will test the relevance of relaxing the isotropy of the pressure. For this, we present a toy model consisting of a thin shell matching an internal Minkowskian core with an external Schwarzschild solution. We show that the shell can be placed as close to its Schwarzschild radius as desired, making the configuration arbitrarily close to the black hole compactness. However, the price to pay is that the distributional tangential pressures of the shell grow without upper bounds as this limit is approached, while the surface density remains finite. In that sense, the Dominant Energy Condition (DEC) is guaranteed to be violated beyond some compactness. In other words, imposing energy conditions sets a compactness bound in the anisotropic case. We revisit and compare our findings with similar results in the literature presenting bounds for anisotropic configurations.  

In a final section we also provide a detailed discussion on how these toy models capture some of the main physical ingredients that are required to violate the Buchdahl limit, and propose ways in which more realistic models can be built taking these toy models as departure points. 
We also analyze some of the arguments that are sometimes used to discard the possibility that these objects are physically meaningful, e.g., the putative violations of causality that might arise if the DEC is violated. 

The paper is structured as follows. In Section~\ref{Sec:Classic_Results}, we revisit key classical results from the literature. Subsection~\ref{SubSec:Buchdahl_Limit} reviews Buchdahl's theorem as a warm-up exercise, highlighting how the assumptions of isotropic pressure and monotonically-decreasing non-negative density function enter the proof. Readers familiar with the argument can skip this subsection. In Subsection~\ref{SubSec:Bondi_Limit}, we examine Bondi’s analysis, discussing the compactness limits he established for both non-negative and negative densities. In Section~\ref{Sec:Bilayer}, we discuss the toy model that illustrates how to violate the Buchdahl limit relaxing the monotonicity of the density. In Subsec.~\ref{Subsec:PositiveBilayer}, we restrict ourselves to positive densities, showing that the upper bound on compactness found by Bondi can be approached arbitrarily closely. We find that such limit is maximized when the internal core is ``empty" (i.e., vanishing density in the internal layer) and study it in detail in Subsec.~\ref{Subsec:ZeroBilayer}. Furthermore, we compare our model with Bondi's Model I. In Subsec.~\ref{Subsec:NegativeBilayer}, we consider the possibility of a negative inner core density, showing that any kind of negative density inner core would open the possibility of having objects as close to the black hole's compactness as desired, in agreement with Bondi's analyses. In Section~\ref{Sec:AdSStars}, we discuss the AdS star model: a particular case of bilayered star in which the inner core is described by AdS spacetime and the outer layer is reduced to a thin shell.
In Section~\ref{Sec:Anisotropy}, we introduce the model of a shell matching an inner Minkowskian core with an outer Schwarzschild metric. We show that it can get arbitrarily close to the compactness of a black hole, although there is a threshold beyond which energy conditions are violated. We use this result to illustrate  the possibility of violating Buchdahl's limit by considering anisotropic pressures. Finally, we conclude in Section~\ref{Sec:Discussion} with a discussion on the features of potentially realistic configurations that are reproduced by these toy models, and ways in which they could be improved in the future. Appendix~\ref{App:FluidSpheres} discusses the constant density perfect fluid spheres, and Appendix~\ref{App:AnisotropicShell} provides the details of the construction of the AdS stars and the second toy model, i.e., the shell matching the internal Minkowskian or AdS core with the external Schwarzschild metric.

 \paragraph*{\textbf{Notation and conventions.}}
In this article, we use the signature $(-,+,+,+)$  for the spacetime metric, and we work in geometric units with $c=G=1$. Einstein's summation convention is used throughout the work unless otherwise stated. For the connection and curvature tensors we use the conventions of Wald's book~\cite{Wald1984}, which we summarize here for completeness. The covariant derivative of a vector is given by $\nabla_a V^b=\partial_a V^b + \Gamma^{b}{}_{ac} V^c$; the commutator of covariant derivatives is $[\nabla_a, \nabla_b]V^c= -R_{abd}{}^c V^d$ and, finally, the Ricci tensor is obtained as $R_{ab}:=R_{acb}{}^c$. The symbolic computations presented here were performed with the assistance of the software xAct~\cite{xAct}.

\section{Classic results: Buchdahl and Bondi}
\label{Sec:Classic_Results}

\subsection{Buchdahl's limit}
\label{SubSec:Buchdahl_Limit}

We first review the Buchdahl limit in an attempt to present with clarity the steps in its derivation. The purpose of this section is twofold, serving both to settle the notation that will be used during this work, and to pose the problem that we will be studying. Importantly, we will enumerate the hypotheses behind the Buchdahl limit, i.e., the existence in general relativity of an upper bound to the compactness of stars in hydrostatic equilibrium. This subsection can be skipped by readers familiar with Buchdahl's analysis.

Consider the following line element that represents static and spherically-symmetric spacetimes: 
\begin{align}
    ds^2 = - f(r) dt^2 + h(r) dr^2 + r^2 d \Omega^2_2.
    \label{Eq:LineElement2}
\end{align}
For some purposes it is useful to use an alternative set of functions:
\begin{align}
    f(r) = e^{2 \Phi(r)}, ~~~~~~~~ h(r)= \frac{1}{1-\frac{2m(r)}{r}},
    \label{Eq:LineElement1}
\end{align}
where $\Phi(r)$ is the redshift function and $m(r)$ is the Misner-Sharp mass function~\cite{Misner1964,HernandezMisner1966}.

We want to study solutions representing a star composed by the stress-energy tensor (SET) of a perfect fluid, i.e., a tensor that reads:
\begin{align}
    T_{ab} = \rho u_a u_b + p \left( g_{ab} + u_a u_b \right),
    \label{Eq:SET}
\end{align}
where $u^a$ is the vector field representing the velocity of the fluid, $\rho$ is the density of the fluid, and $p$ is the pressure. Staticity requires that the vector $u^a$ is aligned with the timelike Killing vector associated with the staticity of the spacetime, $u^a = e^{-\Phi} \delta^a_{\ t}$. Furthermore, we will restrict our considerations to barotropic fluids, i.e., fluids admitting an equation of state of the form $\rho=\rho(p)$. We focus on solutions that are empty, i.e., with $\rho, p = 0$, for $r>R$, corresponding to a sphere filled with a perfect fluid that matches an exterior vacuum spherically symmetric static solution at $r =R$. In virtue of Birkhoff's theorem, the only exterior solution fulfilling these properties is the Schwarzschild solution, and hence we have that for $r>R$:
\begin{align}\label{Eq:SchwMetric}
    & \Phi(r) = \frac{1}{2} \log \left( 1 -  \frac{2M}{r} \right), \\ 
    & m(r) = M, 
\end{align}
with $M \in \mathbb{R}$.

If we plug the ansatz from Eqs.~\eqref{Eq:LineElement2} and~\eqref{Eq:LineElement1} into Einstein equations, we are led to the following equation that determines $\Phi(r) $:
\begin{align}
    \frac{d\Phi(r)}{dr} = \frac{m(r) + 4 \pi r^3 p}{r \left[r-2m(r)\right]},
    \label{Eq:Redshift}
\end{align}
with $m(r)$ defined in terms of an integral of the energy density given in Eq.~\eqref{Eq:SET},
\begin{align}
    m(r) = 4 \pi \int dr' r^{\prime 2} \rho(r').
    \label{Eq:Mass}
\end{align}
Combining Eq.~\eqref{Eq:Redshift} with the conservation relation for the SET,
\begin{align}
    p'(r)=-\left(\rho+p\right)\Phi',
    \label{Eq:Cons}
\end{align}
we arrive at the Tolman-Oppenheimer-Volkoff (TOV) equation, which is given by
\begin{align}
    \frac{dp}{dr} = - (p+\rho) \frac{m(r) + 4 \pi r^3 p}{r \left[r-2m(r)\right]}.
    \label{Eq:TOV}
\end{align}
Alternatively, this equation can be rewritten using a local compactness function defined in terms of the Misner-Sharp mass function as \mbox{$C(r) = 2 m(r)/r$}:
\begin{align}
    \frac{dp}{dr} = - \frac{(p+\rho)}{2r} ~ \frac{\left[ C(r) + 8 \pi r^2 p \right]}{\left[1-C(r)\right]}.
    \label{Eq:P-TOV}
\end{align}
Expressions~(\ref{Eq:Mass}--\ref{Eq:TOV}), together with the equation of state $\rho(p)$, form a closed system of differential equations to be solved for $\Phi,m,\rho$, and $p$. In order to find a specific solution, we take a central density $\rho(0) = \rho_c$ which, by virtue of the equation of state determines a central pressure $p (0) = p_c$. This allows to integrate the TOV equation together with the equation of state and Eq.~\eqref{Eq:Mass} for $m(r)$. Finally, it is possible to determine the redshift by integrating Eq.~\eqref{Eq:Redshift} (or, equivalently, Eq.~\eqref{Eq:Cons}) and matching the redshift of the surface $r=R$ with its Schwarzschild counterpart, given in Eq.~\eqref{Eq:SchwMetric}. 

Buchdahl's limit sets an upper bound on the total mass $M$ that a star with a given radius $R$ can have. To understand what we need in order to surpass Buchdahl's compactness limit, we will express its underlying assumptions in its most conservative version. This is best done by stating the result in the form of a theorem.

\noindent \textbf{Buchdahl's theorem:} Consider the problem of finding the solution to the TOV equations~(\ref{Eq:Mass}--\ref{Eq:TOV}) assuming a perfect fluid with an equation of state that fulfils the following properties: 
\begin{enumerate}
    \item $f(r)$ is at least a piecewise $\mathcal{C}^2$ function, while $h(r)$ is at least piecewise $\mathcal{C}^1$. In particular, both $f'$ and $h$ are continuous at $r=R$. We integrate the equations from the centre $r=0$ with given pressures and densities $\{\rho_c, p_c\}$ towards the surface $r=R$, defined by the condition $p(R)=0$. From that point on ($r>R$) we match an exterior (vacuum) Scwharzschild solution with $\rho = p =0$, independently of the value of $\rho \left( R^{-}\right)$. 
    
    \item The density is a monotonically decreasing function. The continuity and monotonicity assumptions along with the boundary conditions require that \mbox{$\rho(r) \geq 0$}. 
\end{enumerate}
For any such solution, the compactness satisfies the inequality $C(R) < 8/9$.

\textbf{Proof:} Demanding staticity automatically enforces $2M/R \leq 1$, since $C(R)=1$ will lead directly to a diverging pressure. This limit corresponds to a black hole, and physically it is not possible to have a matter content with an energy-momentum tensor that obeys energy conditions (in particular the dominant energy condition~\cite{Wald1984}) sitting statically onto an event horizon.\footnote{Regular black hole models~\cite{Bardeen1968, Hayward1994,Dymnikova1992} exhibit event horizons coexisting with \textit{anisotropic} fluids, but these models fall outside the scope of this discussion.} 
Leaving this particular case aside, 
it is possible to obtain a lower compactness limit by integrating Einstein equations with $\rho(r) \geq 0 $ and $d \rho/dr \leq 0$ and requiring regularity of the solutions. Using the parametrization of the metric given by Eq.~\eqref{Eq:LineElement2}, Einstein equations read
\begin{align}
    G^{t}_{\ t} & = (rh^2)^{-1} h' + r^{-2} \left( 1 - h^{-1} \right) = 8 \pi \rho, \label{Eq:00component}\\
    G^{r}_{\ r} & = (rfh)^{-1} f' - r^{-2} \left( 1 - h^{-1} \right) = 8 \pi p, \\
    G^{\theta}_{\ \theta} & = \frac{1}{2} (fh)^{-1/2} \frac{d}{dr} \left[ (fh)^{-1/2} f' \right] + \frac{1}{2} (rfh)^{-1} f' - \frac{1}{2} (rh^2)^{-1} h = 8 \pi p. 
\end{align}
Eq.~\eqref{Eq:00component}, associated with the tt component, is precisely condition~\eqref{Eq:Mass}. Moreover, because we have assumed isotropic pressure, we can equate $G_{rr}$ and $G_{\theta  \theta}$. Using Eq.~\eqref{Eq:LineElement1} and after some manipulations, we find the following identity:
\begin{align}\label{Eq: equality Gs isotropic}
    \frac{d}{dr} \left[ r^{-1} h^{-1/2} \frac{df^{1/2}}{dr} \right] = (fh)^{1/2} \frac{d}{dr} \left( \frac{m(r)}{r^3} \right). 
\end{align}
We now use the second assumption: that $\rho$ is monotonically decreasing. This implies that the right-hand side of Eq.~\eqref{Eq: equality Gs isotropic} needs to be non-positive, since it is proportional to the derivative of the average density, and the average density is required to be also a monotonically decreasing function. Of course, this implies that the left-hand side needs to be non-positive as well, hence
\begin{align}
    \frac{d}{dr} \left[ r^{-1} h^{-1/2} \frac{d f^{1/2}}{dr} \right] \leq 0. 
\end{align}
Integrating this inequality from the surface of the star located at a radius $R$ to some smaller radius $r$, we find the inequality
\begin{align}
    \frac{1}{rh^{1/2}(r)} \frac{df^{1/2}}{dr} \geq \frac{1}{Rh^{1/2}(R)} \frac{df^{1/2}}{dr}(R) = \left. \frac{(1-2M/R)^{1/2}}{R} \frac{d}{dr} \left( 1 - \frac{2M}{r} \right)^{1/2} \right\rvert_{r = R} = \frac{M}{R^3},
    \label{Eq:Matching}
\end{align}
where we are using the continuity of $f'$ and $h$ to match their values at the surface $r=R$ with their known Schwarzschild counterparts. We now multiply the equation by $r h^{1/2}$ and integrate inwards from the surface $r=R$ to the centre at $r= 0$. We use the value of $f$ at the surface and the expression of $h$ in terms of $m(r)$ to find
\begin{align}
    f^{1/2} (0) \leq \left( 1 - 2M/R \right)^{1/2} - \frac{M}{R^3} \int^R_0 dr \frac{r}{ \sqrt{ 1 - \frac{2 m(r)}{r}}}. 
    \label{Eq:fbound}
\end{align}
The monotonicity condition on $\rho(r)$ implies that $m(r)$ can be, in the best scenario, as small as the value it would have for a uniform density function, that is,
\begin{align}
    m(r) \geq Mr^3/R^3. 
    \label{Eq:}
\end{align}
Thus, the best upper bound that we can provide for $f^{1/2}(0)$ from Eq.~\eqref{Eq:fbound} corresponds to the case when $m(r) = Mr^3 /R^3$, i.e., the constant density scenario yields the optimal bound. Performing the integral for this $m(r)$ we find
\begin{align}
    f^{1/2} (0) \leq \frac{3}{2} \left( 1 - 2 M/R \right)^{1/2} - \frac{1}{2}.
\end{align}
But for regular stellar configurations $f$ must be positive, so we are left with the bound
\begin{align}
    (1 - 2M/R)^{1/2} > 1/3,
\end{align}
and hence
\begin{align}
    C(R)=\frac{2M}{R} < 8 /9, 
\end{align}
which is the so-called Buchdahl limit. 

\emph{Observation 1:} Notice that, by the Tolman-Oppenheimmer-Volkoff equation, it is interchangable to assume $\{\rho \geq 0, d\rho/dr \leq 0 \}$ and $\{p\geq 0, dp/d \rho \geq 0\}$. In that sense, imposing that the pressure and density are positive, together with the causality condition $dp/d \rho \geq 0$, ensure that the density is monotonically decreasing toward the surface of the star. 

\emph{Observation 2:} Notice that we are starting with a perfect fluid and hence the distribution of pressures is isotropic. As we illustrate below, the anisotropic case in which pressures in the angular directions are allowed to grow without bounds does not lead to any limit on the compactness of the object. By forcing the fluid to satisfy energy conditions, though, it becomes possible to find some compactness bounds as well~\cite{Lindblom1984,Ivanov2002,Barraco2003,Boehmer2006,Andreasson2007,Urbano2018} (we will revisit this in Section~\ref{Sec:Anisotropy}).

\emph{Observation 3:} Notice that the proof also tells us which density profile saturates the bound: the uniform density profile. Actually, instead of putting an upper bound to the mass to radius ratio, we can think of it as setting a limit to the maximum mass that a sphere of a given uniform density can display. The properties of this solution are discussed in detail in Appendix~\ref{App:FluidSpheres}. One interesting feature of these constant density solutions is that, for compact enough configurations, the dominant energy condition is violated. This is the case for values of the compactness $2M/R > 6/8$, where we have that $p_c > \rho_0$. This will become relevant later on when we analyze the anisotropic toy model in Section~\ref{Sec:Anisotropy}.

\subsection{Bondi's Limit}
\label{SubSec:Bondi_Limit}

In this section we review Bondi's analysis of compactness bounds for isotropic stars~\cite{Bondi1964}. We refer the reader to~\cite{Karageorgis2007} for an ulterior analysis closing some technical details in Bondi's proof. The framework of Bondi’s analysis is the one presented at the beginning of Section~\ref{SubSec:Buchdahl_Limit}. However, its key novelty lies in relaxing the assumption that the density profile must be monotonically decreasing.

Bondi's analysis begins by introducing the variables $u(r) = m(r)/r$ and $v(r) = 4 \pi r^2 p (r)$. In terms of these variables, the system reduces to:
\begin{equation}\label{Eq:uveqs}
    \frac{1}{r}  \frac{d r}{du} = \frac{(1-2u) \frac{dv}{du} + u + v}{2v - (u^2 + 6uv+v^2)},\quad  4 \pi r^2 \rho = u \frac{dv/du-\beta}{dv/du-\alpha}, 
    \end{equation}
    with 
    \begin{equation}
    \alpha = -\frac{u+v}{1-2u}, \qquad \beta = - \frac{v}{u} \frac{2 - 5u - v}{1 - 2u }.
\end{equation} 
Written in the $(u,v)$ variables, stellar solutions become trajectories in a $(u,v)$ diagram. From those trajectories, it is possible to obtain the physical variables immediately from Eq.~\eqref{Eq:uveqs}, together with the equation for the redshift $\Phi$,
\begin{align}
    \frac{1}{2} \,r\,\Phi' = \frac{u+v}{1-2u}.
\end{align}
Constant compactness curves, those described by $ du/dr = 0$, correspond in these variables to the hyperbolas
\begin{align}
    H \equiv 2v - (u^2 + 6uv+v^2)=0 . 
\end{align}
Also, the integrals of $dv/du = \alpha$ are important and correspond to curves along which $dr/du = 0$, i.e., there is a change in the compactness but the radius of the spheres is kept constant. They are described by the family of parabolas parametrized by $A$:
\begin{align}\label{Eq:PACurves}
    P_A \equiv v = \sqrt{A (1-2u)} - 1 + u. 
\end{align}
Bondi identifies these parabolas with solutions describing thin shells since stellar solutions approaching these parabolas describe narrow matter layers with large densities.

\begin{figure}
    \centering
    \includegraphics[width=0.75\linewidth]{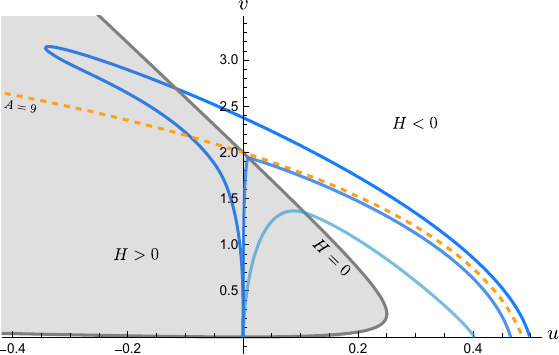}
    \caption{$(u,v)$ diagram of the kind used in Bondi's proof~\cite{Bondi1964}. Regular stellar solutions in the $u>0$ half-plane (with positive $m(r)$) describe curves that start from $(0,0)$ inside the $H>0$ region and increase until they cross $H=0$. After that, while moving within the $H<0$ region, they decrease, always with a more negative slope than $P_{A}$ curves, until they cross $v=0$. These solutions are always bound to take values below the dashed orange $P_{\rm A}$-curve with $A=9$, hence they must have a compactness $C(R)<C_{\rm B}$. Solutions in the $u<0$ half-plane (with negative $m(r)$) start from $(0,0)$ and can take negative $u$ values as large as desired. Then they have a turning point, cross $H=0$, and keep decreasing until they cross the $v=0$ axis. These solutions can have any compactness $C(R)<1$. We have drawn three representative curves meant to describe stars with increasing compactness. }
    \label{Fig:uvDiagram}
\end{figure}

Now, we need to consider how solutions describing a stellar interior behave in these $(u,v)$ variables. If we restrict to non-negative values of the density, $u$ needs to be non-negative as well. Furthermore, Bondi also shows that non-negative densities enforce non-negative pressures. Thus, an interior with non-negative densities is confined to lie within the first quadrant of the $(u,v)$ diagram in Fig.~\ref{Fig:uvDiagram}. In this figure we have drawn, in light blue, two curves representing stars with non-negative densities.
In the $(u,v)$ diagram, a generic solution describing a stellar interior with finite central pressures (for example, the blue curves in Fig.~\ref{Fig:uvDiagram}) needs to be such that it starts from the origin $(0,0)$ and moves within region $H>0$, intersecting $P_{A}$ curves of increasing $A$ until it hits the hyperbola $H=0$ with exactly the same slope as the $P_{A}$ that is intersecting it at the crossing point (to ensure the finiteness of~\eqref{Eq:uveqs}).

Then, the solution enters the region $H<0$ and acquires negative slope, intersecting $P_{A}$ curves with decreasing $A$, eventually reaching $v=0$ at a given value of $u$ that fixes the compactness of the star. Now, we attempt to construct the most compact solution possible within the positive $(u,v)$ region. Restricting ourselves to $u \geq 0$, the highest point at which it is possible to cross $H = 0$ is $v = 2$. Once in the $H<0$ region, the curve that optimizes the compactness is the $A=9$ parabola, which intersects $v=0$ at $u = 6 \sqrt{2} -8$. Since a regular solution must lie below both these curves, this proves the existence of an upper bound for the compactness of stars with non-negative densities, namely 
\begin{equation}\label{Eq:BondiBound}
    C(R)<C_{\rm B}\equiv 12 \sqrt{2} - 16.
\end{equation}
The thick blue line in Fig.~\ref{Fig:uvDiagram} represents a generic configuration near but below Bondi's limit.

Allowing solutions to take values in the $u<0, v>0$ region, it is possible to obtain curves that grow in $v$ towards negative $u$ values and which are then matched with curves arbitrarily close to $P_{A}$ parabolas with $A$ values as large as desired. Since the $P_{A}$ curves cross $v = 0$ at the positive $u$ values
\begin{equation}
    u_{+} = 1-A+\sqrt{A\left(A-1\right)},
\end{equation}
we see that
\begin{equation}
    \lim_{A\to\infty}u_{+}=\frac{1}{2}-\frac{1}{8A}+\order{\frac{1}{A^{2}}},
\end{equation}
and hence we have that $C(R)<1$ and the black hole limit can be approached as much as desired. 
The darkest blue line in Fig.~\ref{Fig:uvDiagram} represents a generic configuration with a negative energy core and a compactness beyond Buchdahl's and Bondi's bounds.

After presenting this general framework, Bondi proposed two specific stellar models, Model I and Model II. These models serve, respectively, to illustrate how to saturate the Bondi bound and how to construct configurations that surpass it thanks to negative energy cores. These models incorporate thin shells in their construction. At first glance, thin shells seem to require anisotropic pressures, which would be against the starting isotropy assumption. However, here we will show explicitly that these anisotropies do not arise since the pressures identically vanish due to a precise fine-tuning of the parameters of the models. We present a set of bilayered models with no distributional matter in the following sections as a complement to Bondi's models. These models further illustrate how Buchdahl's and Bondi's limits can be approached and surpassed with specific examples.

\section{Bilayered stars with non-monotonically decreasing density profiles}
\label{Sec:Bilayer}
\begin{figure}[H]
    \centering
    \includegraphics[width= 0.35\textwidth, height= 0.35\textwidth]{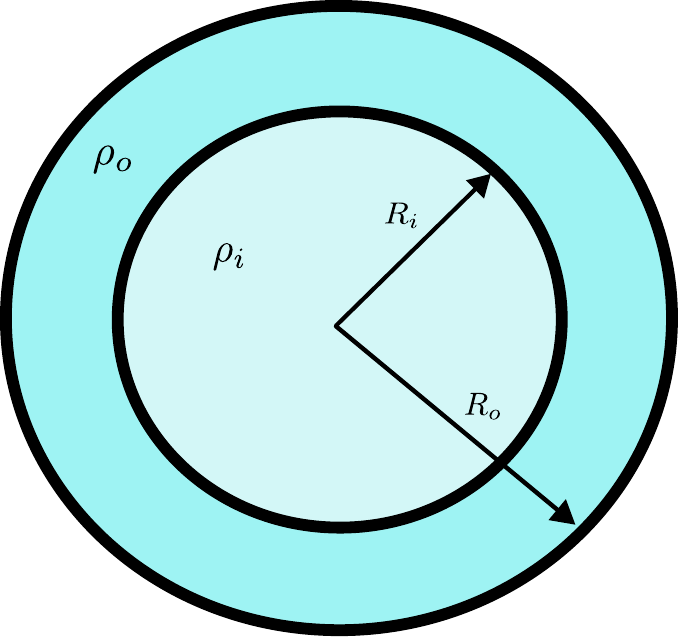}
    \caption{Pictorial representation of the two-layered toy model that we are considering in this section. The density of the outer layer is bigger than that of the inner layer, $\rho_i < \rho_o$.}
    \label{Fig:Bilayer}
\end{figure}

In this section, we will focus on the study of a toy model displaying a very simple outward increasing density profile. Specifically, we consider a star whose internal structure displays two constant-density layers: 
\begin{align}
    \rho(r) = \begin{cases}
      \rho_i  & r < R_i \\
      \rho_o & R_i < r < R \\
      0 & r> R
    \end{cases},
\end{align}
with $\rho_i < \rho_o$ and $R_i < R$. As indicated in previous sections, the exterior metric ($r > R$) is Schwarzschild. The setup is represented in Fig.~\ref{Fig:Bilayer}.
We impose that the Misner-Sharp mass function is continuous everywhere. Physically, this means that all the mass of the system is entirely provided by the two layers (i.e. there are not additional shell-like contributions). This implies that we have
\begin{align}
    m(r) = \begin{cases}
      \frac{4}{3} \pi r^{3} \rho_{i}  & r < R_i \\
      \frac{4}{3} \pi R_i^3 \rho_{i} + \frac{4}{3} \pi \rho_{o} (r^3 - R_i^3) & R_i < r <R \\
      M & r> R
    \end{cases},
    \label{Eq:MisnerSharpBi}
\end{align}
together with the constraint:
\begin{align}
   M =  \frac{4}{3} \pi R_i^3 \rho_{i} + \frac{4}{3} \pi \rho_{o} (R^3 - R_i^3).
   \label{Eq:Constraint}
\end{align}
As mentioned before, $M$, which represents the ADM mass of the spacetime, cannot surpass the black hole limit: $M<R/2$. The inner radius $R_i$ can, at most, range between the centre of the star and the total radius
\begin{align}
    0 < R_i < R,
\end{align}
since the star would become single-layered in both limits. The outer density $\rho_o$ is allowed to vary within the range
\begin{align}
    \rho_c <\rho_{o}<\infty,
\end{align}
with $\rho_{c}=3M/4\pi R^3$ being the critical density, which corresponds to the limiting case in which $\rho_i = \rho_o$, i.e., the star has constant density, and therefore it cannot be more compact than what the Buchdahl limit dictates.

Bilayered fluid stars satisfy Israel junction conditions at $R_{i}$ since, by virtue of~\eqref{Eq:Cons}, the discrete jump in $\rho$ between the inner and outer layers translates into a compensatory discontinuity in $p'$ that guarantees the continuity of $\Phi'$. Consequently, the metric components $g_{tt}$ and $g_{rr}$ are piecewise $\mathcal{C}^2$ and $\mathcal{C}^{1}$ functions, respectively. This discontinuous behavior in the $tt$ component of the stress-energy tensor is of the same kind as the one that occurs at the surface of the star and, therefore, does not carry along the introduction of distributional stress-energy sources of any kind.

The structure of this bilayered fluid sphere is specified by the set of five parameters $\{\rho_{i},R_{i},\rho_{o},R,M\}$. However, not all of the parameters are independent. First of all, since the TOV equation that we are going to solve does not incorporate any additional length scales (such as the one that appears in the semiclassical TOV~\cite{Carballo-Rubio2017,Arrecheaetal2021}), for the numerical problem we can fix one of the parameters as a unit to which the rest are compared. In our case, we will choose $R$ to be that parameter of choice: for all plots, we set $R = 1$.  Moreover, we have the constraint given in Eq.~\eqref{Eq:Constraint}, which reduces the number of independent parameters to four. We will rewrite the parameter $\rho_i$ in terms of the other variables
\begin{align}
    \rho_{i}=  \frac{3 M}{4\pi R_{i}^3} +\rho_{o}\left(1 - \frac{R^3}{R_{i}^3} \right), 
\label{Eq:RhoI}
\end{align}
and take the independent parameters to be \mbox{$\{ M, R_i, \rho_o \}$.}

We decide to integrate the TOV equations from the surface of the star towards its centre. Although this is entirely equivalent to outward integrations from a regular centre, our strategy allows to select the compactness of the fluid sphere beforehand, thus proving to be more efficient to search for compactness bounds from the numerical point of view.

Solutions for the functions $p$ and $\Phi$ at the inner layer are straightforward to obtain by integrating the TOV equation (see Appendix~\ref{App:FluidSpheres}), and fixing the corresponding integration constants in terms of the parameters of the outer layer. 
Defining the constants $\Phi_{\text{i}}$ and $p_{i}$ as
\begin{equation}
\Phi_{i}= \Phi(R_i),\quad p_{i}= p(R_i),
\end{equation}
we obtain
\begin{equation}
  \Phi(r)=\Phi_{i}\log\left[3\left(p_{i}+\rho_{i}\right)-\left(3p_{i}+\rho_{i}\right)\sqrt{\frac{3-8\pi r^2\rho_{i}}{3-8\pi R_{i}^2\rho_{i}}}~\right]-\Phi_{i}\log\left(2\rho_{i}\right), \label{Eq:InnerPhi}
\end{equation}
and
\begin{equation}
    p(r)=-\rho+e^{-\Phi(r)+ \Phi_i}\left[p_{i}+\rho_{i}\right],  \label{Eq:InnerP}
\end{equation}
for $r < R_i$, i.e., inside the inner layer.
For the outer layer, instead, the solution can be written as combinations of elliptic functions of cubic roots~\cite{Wyman1939}, resulting in lengthy and non-illuminating expressions that we omit here but which were obtained and evaluated numerically using the software \textit{Mathematica.} We have attached a Mathematica notebook containing the pressure, mass, and redshift functions of the outer layers. The query about the maximum compactness allowed by these bilayered toy models reduces to an inquiry about whether there exists a maximum possible value for the total mass $M$ (always below the black hole limit, or $M < R/2$), beyond which there is not any solution in the family preserving a finite and regular pressure  function everywhere inside (and equivalently for the redshift function). 

We distinguish two situations: the case in which the inner density $\rho_i$ is strictly positive, which we explore in detail in Subsec.~\ref{Subsec:PositiveBilayer}, with the limiting case $\rho_i = 0$ presented separately in Subsec.~\ref{Subsec:ZeroBilayer}; and the case in which the density of the inner layer is negative, which we discuss in Subsec.~\ref{Subsec:NegativeBilayer}. For positive inner densities, the Misner-Sharp mass function remains positive throughout the configuration, and we demonstrate that the maximum compactness bound found by Bondi, Eq.~\eqref{Eq:BondiBound}, can be approached arbitrarily. For negative inner densities, regions with negative Misner-Sharp mass appear in the interior, allowing for configurations that reach arbitrarily high compactness, all the way up to the black hole limit, in full agreement with Bondi's conclusions. Unlike Bondi's Model I and II, the bilayer model presented here gives rise to a well-defined geometry throughout.

\subsection{Bilayered stars with positive core densities}
\label{Subsec:PositiveBilayer}

\begin{figure}
    \centering
    \includegraphics[width=0.7\linewidth]{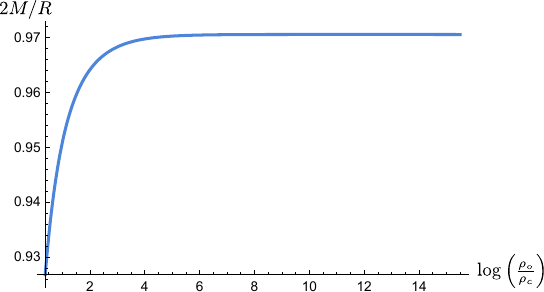}
    \caption{Maximum compactness limit as a function of the quotient between the energy density of the outer layer and the critical density for stars with $\rho_{i}>0$.  For $\rho_{o}/\rho_{c}\to1$, the compactness limit tends to the Buchdahl limit whereas for $\rho_{o}/\rho_{c}\to\infty$ it approaches the Bondi limit.}
    \label{Fig:MaxCP}
\end{figure}

\begin{figure}
    \centering
    \includegraphics[width=0.75\linewidth]{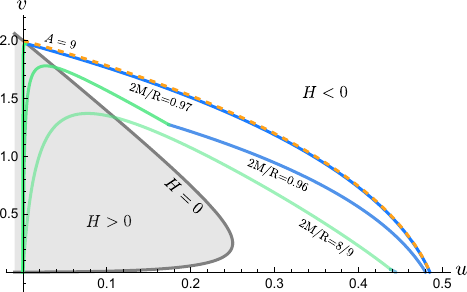} 
    \caption{$(u,v)$ diagram of bilayered stars with non-negative energy densities. 
    We have represented the $v(u)$ curves from the inner layer in green and those from the outer layer in blue. As the compactness approaches the Bondi compactness, the bilayered model acquires characteristics very similar to Bondi's model I, namely an inner layer that lies on the vertical axis and an outer layer that approaches the $P_{A}$-curve with $A=9$, which is the dashed orange line in the figure. Green curves closer to the vertical axis correspond to smaller inner-layer densities.}
    \label{Fig:UVPos}
\end{figure}

Let us begin analyzing the case in which we have $\rho_i > 0$. The positivity of the right hand side of Eq.~\eqref{Eq:RhoI} imposes a lower bound on $R_{i}$. Explicitly, we have that
\begin{align}
    \left(1-\rho_{c}/\rho_{o}\right)^{1/3} R < R_{i} < R.
    \label{Eq:RadI}
\end{align}
We demonstrate that there is an upper bound to the compactness that these stars can achieve, beyond which no regular solutions exist. To establish this, we fix the parameters $R_{i}$ and $\rho_{o}$, and we determine numerically the value of $M_{\infty}(R_{i}, \rho_{o})$ that leads to a solution where the central pressure becomes infinite. This solution defines the maximum mass that a sphere with outer density $\rho_{o}$ and inner radius $R_{i}$ can sustain. Spheres with masses below this maximum will have finite pressures throughout, whereas for larger masses, an ``infinite pressure surface'' emerges at some finite radius of the star $r > 0$.
\begin{figure}
    \centering
    \includegraphics[width=0.75\linewidth]{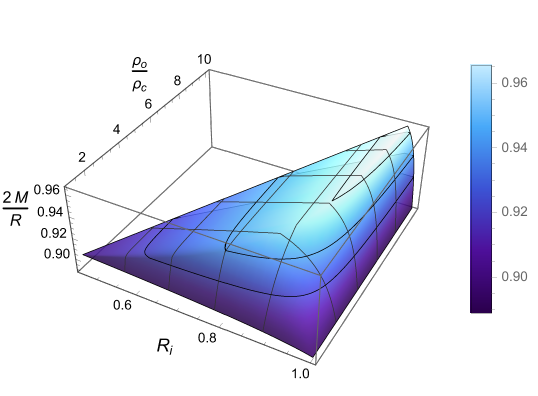}
    \caption{Three dimensional representation of the surface in the $\{R_{i},\rho_{o},M\}$ space corresponding to solutions with an infinite central pressure and  $\rho_{i}>0$. Every point on this surface lies above the standard Buchdahl limit  $2M/R=8/9$, which is reached in the $\rho_{o}/\rho_{c}\to1,~R_{i}/R\to1$ limits. Lighter colours denote larger values of $2M/R$, and the black contour lines correspond to the \mbox{$2M/R=\{0.92,0.94,0.96\}$} planes. 
    This surface can be extended further in the $\rho_{o}/\rho_{c}\gg 1$ direction where we find an asymptotic approach to the maximum compactness value of $2M/R\approx 0.9706$. We have included a Mathematica notebook containing the pressure, mass, and redshift functions of both the inner and outer layers, as well as this 3D plot, so that any interested reader can visualize it from other angles.}
    \label{Fig:BilayeredRhoP}
\end{figure}
By varying the values of $\rho_o$ and $R_i$, we can map out a surface in the three-dimensional parameter space $\{R_{i}, \rho_{o}, M\}$ that precisely represents the value of $M_{\infty}(R_i, \rho_o)$ for each pair $\left(R_i,\rho_o\right)$. This surface is shown in Fig.~\ref{Fig:BilayeredRhoP}. The vertical axis represents the maximum mass allowed for a given $R_i$ and $\rho_o$. All points on this surface lie between two constant $2M/R$ planes: one above Buchdahl's limit ($2M/R = 8/9$), which it intersects in the limits $\rho_{o}/\rho_{c} \to 1$ and $R_{i}/R \to 1$, and the other below the black hole limit ($2M/R = 1$), which it neither intersects nor approaches asymptotically. Extending Fig.~\ref{Fig:BilayeredRhoP} in the direction of $\rho_{o}/\rho_{c} \gg 1$, we find that $C(R)$ saturates at a maximum value which approximately is
\begin{align}\label{Eq:CompLimRhoP}
    C_{\text{max}} = 2\sup_{ R_{i},\rho_o } \frac{M_{\infty} (R_i,\rho_o)}{R} \approx 0.9706.
\end{align}
To the level of our numerical precision it agrees with Bondi bound~\eqref{Eq:BondiBound}. In fact, we find that there is no gap between this uppermost bound~\eqref{Eq:CompLimRhoP} and the Bondi bound. Figure~\ref{Fig:MaxCP} shows how this maximum compactness limit scales with the quotient between the outer and critical densities. Finally, in Figure~\ref{Fig:UVPos} we have plotted three solutions belonging to this family with positive, albeit small, inner-layer densities.

\subsection{Bilayered stars with vanishing core densities}
\label{Subsec:ZeroBilayer}
A particularly illustrative case arises when the density of the inner core is zero, a feature also present in Model I proposed by Bondi. We treat it separately because its simplicity allows to find the compactness limit in an easier way. Moreover, we can provide numerical evidence for the absence of any finite gap between the maximum compactness attained by this bilayered model and Bondi's bound. In the present situation, the Misner-Sharp mass vanishes everywhere inside the core and, for a fixed value of the density $\rho_o$ and the mass $M$, Eq.~\eqref{Eq:RhoI} enforces the radius $R_i$ to be precisely
\begin{equation}\label{Eq:Rirho0}
    R_{i}=\left(1-\frac{\rho_{c}}{\rho_{o}}\right)^{1/3}R,
\end{equation}
where we recall that the expression of $\rho_c$ in terms of $\{M,R\}$ is $\rho_c = 3M/4 \pi R^3$. This constraint also restricts the pressure to be a positive, inwards-increasing monotonic function. This is because, starting from a positive pressure that grows inward, the numerator of the TOV equation~\eqref{Eq:TOV} is only allowed to change sign if $m(r)<0$. Actually, we can explicitly integrate the TOV equation in the inner layer and find the following pressure profile:
\begin{equation}
    p=\frac{1}{2\pi r^2+\kappa},
\label{Eq:PressureZero}
\end{equation}
where $\kappa$ is an arbitrary integration constant, corresponding to the inverse of the pressure at the origin, whose value is obtained from the matching with the outer layer, i.e., 
\begin{equation}\label{Eq:KappaValue}
    \kappa=\frac{1}{p_{i}}-2\pi R_{i}^2=\frac{1}{p_{i}}-2\pi R^2\left(1-\frac{\rho_{c}}{\rho_{o}}\right)^{2/3}.
\end{equation}
In Subsection~\ref{Subsec:PositiveBilayer} we noticed that the regular bilayered star with positive density that exhibits the largest possible compactness $C_{\text{max}}$ would be obtained in the limit of a thin and infinitely dense shell surrounding an interior of (nearly) vanishing mass. In the present case, where the inner core has exactly zero mass, the  maximum compactness is attained by the solution~\eqref{Eq:PressureZero} with $\kappa=0$, precisely the one that makes the pressure profile diverge exactly at $r=0$. To find this compactness bound, we  calculate how $p_{i}$ scales in the $\rho_{o}/\rho_{c}\to\infty$ limit (the limit in which the outer shell is arbitrarily thin) for different values of the compactness. 

In our numerical explorations, we see that $p_i$ always approaches a constant positive value, which we denote as $P_{i}^{~(\text{max})}$, as we vary $\rho_o$ while keeping the mass fixed for $\rho_{o}\gg \rho_{c}$.  By calculating $P_{i}^{~(\text{max})}$ for a range of compactness values between the Buchdahl and black hole limits, we find that, if the density of the outer layer is made very large, the value $\kappa=0$ is found very close to the Bondi limit, defined in Eq.~\eqref{Eq:BondiBound}. In particular, we find the scaling
\begin{equation}\label{Eq:LimitScaling}
    C(R)\approx C_{\rm{B}}-\frac{\alpha}{R^{2}\rho_{o}}, 
\end{equation}
where $\alpha$ is a dimensionless positive constant. The Bondi bound is only reached in the singular limit $\rho_{o}\to\infty$, which would correspond to putting all the mass of the star inside a thin shell---since, in virtue of~\eqref{Eq:Rirho0}, it implies that $R_{i}\to R$. Although the solution with $\kappa = 0$ is, strictly speaking, singular, we can find solutions whose compactness approaches Bondi's bound as much as desired through a physical model that is purely isotropic and can be appropriately matched with a Schwarzschild exterior. 

\paragraph*{\textbf{Comparison with Bondi's Model I.}}

In Figure~\ref{Fig:UVPos}, we have plotted two configurations in the $(u,v)$ variables belonging to our family of vanishing core densities. As the external crust becomes thinner, the compactness approaches Bondi's bound more closely. The portion of the curve with a negative slope, where $v$ decreases as $u$ increases, corresponds to the thick shell region. Regardless of the shell thickness, these curves never coincide with the parabolic curves defined by a fixed parameter $A$. Instead, they consistently move downward, intersecting parabolas with different values of $A$, which is the crucial difference with respect to Bondi's Model I.

Model I is described as a vanishing energy core with non-vanishing pressure connecting through a thin envelope to an external Schwarzschild geometry. The vanishing energy core is represented by the curve $ u = 0$ and $v \in [0, 2 - \epsilon]$, $\epsilon >0$, where
\begin{align}
    \rho = 0, \qquad p = \frac{p_c}{1+2\pi p_c r^2}, \qquad 2 \pi p_c r^2 = \frac{v}{2-v}, \qquad 0 \leq r \leq \sqrt{\frac{2 - \epsilon}{2 \pi p_c \epsilon}},
\end{align}
with $p_c$ being the value of the pressure at the centre $r=0$. The envelope is represented by $v = ( 3 - \epsilon) (1 - 2u)^{1/2} - (1-u)$ and $u \in [0, u_s]$, with 
\begin{align}\label{Eq:uBound}
    u_s = \left( 3 - \epsilon \right) (8 -6 \epsilon + \epsilon^2)^{\frac{1}{2}} - \left( 8 - 6 \epsilon + \epsilon^2 \right) = 6 \sqrt{2} - 8 - \order{\epsilon}.
\end{align}
As the envelope is strictly following one of the parabolic curves, it represents from the start an infinitely-thin limit. This is so because along such curve $r$ does not vary, and the energy density becomes infinite (cf. Eqs.~\eqref{Eq:uveqs}). By making $\epsilon$ arbitrarily small, the model can approach the Bondi compactness limit as closely as desired.

To better understand Bondi's Model I, let us perform a detailed thin-shell analysis following Israel's formalism: 
Take a spherical shell with radius $R$, such that its proper area is $4 \pi R^2$. Let the spacetime be described by a external Schwarzschild patch that we denote with a $+$ sign and the internal patch that we denote with a $-$ sign. Generically, the metric reads:
\begin{align}
    & ds^2_{+} = - f_{+}(r_{+}) dt_{+}^2 + h_{+}(r_{+}) dr_{+}^2 + r_{+}^2 d \Omega_2 ^2, \\
    & ds^2_{-} = - f_{-}(r_{-}) dt_{-}^2 + h_{-}(r_{-}) dr_{-}^2 + r_{-}^2 d \Omega_2 ^2.
    \label{Eq:thinshellbulks}
\end{align}
The external Schwarzschild patch corresponds to $f_{+}(r_{+}) = 1 - 2M/r_{+}$ and $h_{+}(r_{+}) = 1/f(r_+)$.  The internal geometry has a vanishing mass function and, as such, $h(r_{-})=1$ and the redshift function is given by  
\begin{equation}
    f(r_{-})=( 2\pi r_{-}^2+1/p_{c})^2.    
\end{equation}
We locate the thin shell at $r_{+} = r_{-} = R$. The matching is worked out in detail in Appendix~\ref{App:AnisotropicShell}, and we find that the shell displays a distributional stress-energy tensor with the following densities and pressures:
\begin{align}
    & \rho = \sigma \delta (r - R), \\
    & p_r = 0, \\
    & p_t = \tilde{p}_t \delta(r -R),
\end{align}
with $\sigma$ and $\tilde{p}_t$ given by
\begin{align}
     \sigma & = \frac{1}{4 \pi R} \left(1 - \sqrt{1-\frac{2M}{R}} \right), \nonumber \\
     \tilde{p}_t & = \frac{1}{8  \pi R } \left(\sqrt{1-\frac{2M}{R}} -1 \right) +\frac{1}{16  \pi R} \left(\frac{2M}{R} \frac{1}{\sqrt{1-\frac{2M}{R}}} - \frac{8\pi p_c R^2}{(1 + 2 \pi p_c R^2)}\right).
\end{align}
Now, for a given value of the compactness, there is always a specific value of $p_c$ such that the tangential pressure vanishes, i.e., $\tilde{p}_t=0$, which corresponds to
\begin{align}
 p_c= -\frac{R \left(\sqrt{1-\frac{2 M}{R}}-1\right)+M}{2 \pi  R^2 \left(R \left(3 \sqrt{1-\frac{2 M}{R}}-1\right)+M\right)}.
\end{align}
In this scenario, the shell is held in place by the pressure from the matter content of the inner region, despite the shell itself being pressureless. This situation is peculiar because it involves an interior composed of unusual matter that exerts pressure without possessing any energy density. We discover that the compactness cannot be increased to the black hole limit, as there is a specific value at which the solution becomes singular for $r=0$, since the pressure there becomes infinite, i.e., $p_c \to \infty$. Increasing the compactness only moves this singularity to finite radii. By looking at Eq.~(\ref{Eq:Densities}), one can check that  the threshold value of $p_c = +\infty$ corresponds precisely to a compactness

\begin{align}
    C(R) = 12 \sqrt{2} - 16,
\end{align}
which is precisely Bondi's bound $C(R) = C_B$. 
Thus, we have seen that Bondi's Model I corresponds to the distributional limit of our models, with each providing further insights into the other.
In fact, for any given compactness $C_1(R)$ lying between Buchdahl’s and Bondi's limits (i.e., $C_{\rm Buchdahl}<C_1(R)<C_{\rm Bondi}$), there is always a bilayered model with that compactness, as well as vanishing density cores and sufficiently thin crusts which ensure that the integration of the TOV equation from the surface inwards leads to a completely regular configuration.

\subsection{Bilayered stars with negative densities in the core}
\label{Subsec:NegativeBilayer}

We have demonstrated that an outward-increasing, non-negative density profile allows for compactness exceeding the Buchdahl limit, as established by Bondi, and we have constructed a toy model that nearly saturates this bound. Bondi further argues that permitting negative inner densities $\rho_{i}$ removes this restriction, allowing for configurations approaching the black hole compactness limit. In this Subsection, we provide an explicit example within the bilayered model, showing that for negative of $\rho_i$, objects can be constructed arbitrarily close to this limit. However, this relaxation also permits violations of the weak (and consequently the dominant) energy conditions.

In the case examined previously, we found that the solution which maximizes the compactness looks (approximately) like a star with all its positive mass stored in a very narrow outer layer. The inner layer, on the contrary, occupies the whole bulk of the star and does not contribute to the mass at all.  Such arrangement of the two layers is optimal in the sense that it maximizes the mass of the configuration. Instead, allowing for a negative density core enables the construction of stars for which the configuration that minimizes the central pressure does not require a very narrow and highly dense outer layer. Examples of these solutions are found by selecting the position of $R_{i}$ in such a way that the pressure remains almost constant inside the core. The compactness of these solutions can be arbitrarily close to the black hole limit. The rest of this subsection is devoted to prove this statement in detail. 

Assume a bilayered model with an internal layer of negative density $\rho_{i}$. In order to proceed with the proof, we need to take a step back and consider for a moment the uniform-density, single-layer model ($R_{i}=0$). By inspection of Eqs.~\eqref{Eq:MisnerSharpBi} and~\eqref{Eq:Constraint}, we see that, for $\rho_{o}>\rho_{c}$, the Misner-Sharp mass $m(r)$ takes negative values in the interval $0\leq r<\left(1-\rho_{c}/\rho_{o}\right)^{1/3} R$. Extrapolated to $r=0$, this would point to the existence of a negative delta contribution to the density at the core $r=0$, meaning that the extrapolation of this configuration all the way to the centre of the star is not regular. Here, we only use this irregular configuration to generate the outer layer of the total geometry. In these configurations it is easy to see that the width of the negative-mass region grows with $\rho_{o}$. Now, by the TOV equation~\eqref{Eq:TOV} we see that, at the surface $r=R$, pressure always increases inwards:
\begin{equation}
    p' = - \frac{M\rho_{o}}{\left(R-2M\right)R}<0.
\end{equation}
Moreover, a negative Misner-Sharp mass can produce a maximum in the pressure at some $r=r_{\text{max}}$, which, by the Eq.~\eqref{Eq:TOV}, corresponds with the point when \mbox{$m(r_{\text{max}})=-4\pi r_{\text{max}}^3 p(r_{\text{max}})$}. Values of $\rho_o$ bigger than a critical value $\rho_{\text{sep}}$ (whose specific value can be found numerically) ensure that it is possible to generate configurations with pressure profiles that are everywhere finite, in the same way that it is done in~\cite{Arrecheaetal2021}: starting from $p(R)=0$ at the surface, then reaching a maximum value at $r=r_{\text{max}}$, to finally decrease until the centre $r=0$. In the $r\to0$ limit, we obtain the following behavior:
\begin{equation}
        p\simeq -\rho + k \sqrt{r},\quad k>0.
\end{equation}  
As mentioned above, if only one layer is allowed, these configurations display a curvature singularity at $r=0$. However, if we allow for an inner layer with $\rho_{i}<0$ that provides a physical origin to the negative Misner-Sharp mass, then we can find a range of matching radii $R_{-} \leq R_{i} \leq R_{+}$ that guarantees that $p' \geq 0$ inside the core and exactly zero at $r=0$, thus yielding a finite and well-defined pressure everywhere.
Inside the core, the solutions for the  Misner-Sharp mass, redshift, and pressure are given by Eqs.~(\ref{Eq:MisnerSharpBi}--\ref{Eq:InnerP}). 

\begin{figure}
    \centering\includegraphics[width=0.65\textwidth]{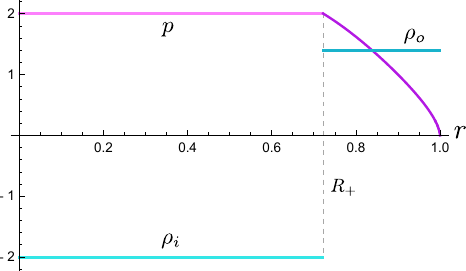}
    \caption{Typical pressure and density profiles for stars displaying a constant-pressure inner layer such that $p+\rho_{i}=0$. The star shown in this Figure has $2M/R=0.99,~R^{2}\rho_{o}=1.4$ and $R_{i}=R_{+}\approx0.72R$. In the limit where the outer layer is infinitesimally thin, these solutions reduce to the AdS star model (see Sec.~\ref{Sec:AdSStars} below).
    It is always possible to find stars belonging to this family for any compactness $2M/R<1$.}
    \label{Fig:RP}
\end{figure}

\begin{figure}
    \centering\includegraphics[width=0.65\textwidth]{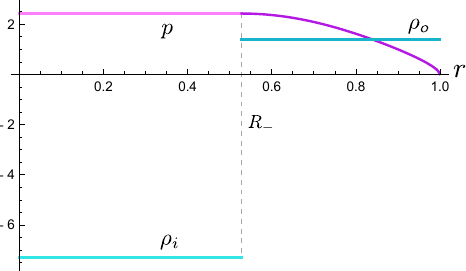}
    \caption{Typical pressure and density profiles for stars displaying a constant-pressure inner layer such that $3p+\rho_{i}=0$. The star shown in this Figure has $2M/R=0.99,~R^{2}\rho_{o}=1.4$ and \mbox{$R_{i}=R_{-}\approx0.53R$}. In the limit where the outer layer is infinitesimally thin, these solutions reduce to the Einstein static star model (see Sec.~\ref{Sec:AdSStars} below).
    It is always possible to find stars belonging to this family for any compactness $2M/R<1$.}
    \label{Fig:RM}
\end{figure}

In fact, the two boundaries of this interval $R_{-}$ and $R_{+}$ correspond to the choices of internal radii that make $p_i = - \rho_i/3$ and $p_i = - \rho_i$ respectively. These special values will be analyzed in detail in Sec.~\ref{Sec:AdSStars}, as well as their connection with ultracompact objects found through semiclassical analyses. 
On the one hand, for $R_i = R_{+}$, the inner core has constant (positive) pressure in a way that guarantees $p+\rho_{i}=0$ everywhere inside the core. The redshift function adopts the simple form
\begin{equation}\label{Eq:Phi_i}
\Phi(r)=\Phi_{i}\log\left(\sqrt{\frac{3-8\pi r^2\rho_{i}}{3-8\pi R_{i}^2\rho_{i}}}\right).
\end{equation}
It is straightforward to check that this solution is regular in the range $0\leq r<R_{i}$, since the only divergences in~\eqref{Eq:Phi_i} either come from
\begin{equation}
    8\pi r^2 \rho_{i}-3=0,\quad \text{or}\quad 8\pi R_{\text{i}}^2 \rho_{i}-3=0,
\end{equation}
which do not have real roots if $\rho_{i}<0$. On the other hand, the choice $R_i = R_{-}$ leads to an inner core that is characterized by a constant redshift function $\Phi(r)=\Phi_{i}$ and which satisfies the condition $3p+\rho_i=0$. Figures~\ref{Fig:RP} and~\ref{Fig:RM} show plots of these two solutions with constant-pressure inner cores. For any choice of $R_i$ in between those two values, namely $R_{-}<R_{i}<R_{+}$, we obtain regular solutions with $p'>0$ inside the core. Regular solutions can also be found for $R_{i}\gtrsim R_{+}$ and for $R_{i}\lesssim R_{-}$ although we do not focus on them here since our aim is only to show that it is possible to find solutions as compact as desired. 

Let us now prove that for $R_i \in [R_{-},R_{+}]$ the pressure is always bounded in the region $r \in [R_i,R]$ for any value of $\rho_o$ greater than a given value $\rho_{\text{sep}}$. Recall that the density of the inner layer is fixed in terms of the other parameters through Eq.~\eqref{Eq:RhoI}. In the $R_{i}\to0$ limit, it diverges towards negative infinity as
\begin{equation}
    \rho_{i}\propto \left(\rho_{c}-\rho_{o}\right)\frac{R^3}{R_{i}^{3}},
\end{equation}
provided that  $\rho_{o}\geq\rho_{\text{sep}}> \rho_{c}$. Since we have chosen $\rho_{o}$  such that pressure in the outer layer remains bounded regardless of the radius $R_{i}$ of the inner layer, and since by moving $R_{i}$ we can construct inner regions with  $\rho_{i}\in(-\infty,0)$, there is always an interval $R_{i}\in[R_{-},R_{+}]$ for which $p_{i} /\rho_{i}\in\left[-1,-1/3\right]$. Within this interval of the parameter space, the interior solution for the redshift, given by Eq.~\eqref{Eq:InnerPhi}, is regular, and approaches the solution of Eq.~\eqref{Eq:Phi_i} as $R_{i}\to R_{-}$, and the constant solution $\Phi(r)=\Phi_{i}$ as $R_{i}\to R_{+}$. Finally, since  $\rho_{\text{sep}}$ exists for every $2M/R<1$, we conclude that, by violating the assumption $\rho_{i}\geq0$, we can obtain a new family of solutions describing stars with constant-pressure inner layers for which there is no upper compactness bound. 
\begin{figure}
    \centering
    \includegraphics[width=0.7\linewidth]{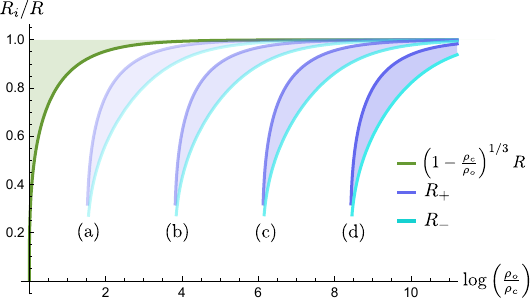}
    \caption{Regions of regular solutions for different compactness values. The shaded green area represents the parameter space for stars with $\rho_i>0$, with its boundary is indicated by the green curve. Within this region we find the maximum compactness value $C_{\text{max}}\approx 0.9706 $. 
    The shaded blue regions (a), (b), (c) and (d) correspond to regions of the space of parameters (bounded by the curves $R_{+}$ and $R_{-}$) for which we find regular solutions with $2M/R=\left\{0.998,0.9998,0.99998,0.999998\right\}$, respectively. These solutions are necessarily in the $\rho_{i}<0$ region and can be found for any compactness value $2M/R<1$.}
    \label{fig:enter-label}
\end{figure}

As a summary, the main consequence of not constraining the sign of $\rho_i$ is that we allow to explore a broader region of the space of parameters $ \{ R_i, \rho_o, M\}$. This is depicted in Fig.~\ref{fig:enter-label} where, for some fixed mass value, only solutions inside the green region are allowed to have $\rho_i>0$. The green curve $R_{i}=\left(1-\frac{\rho_{c}}{\rho_{o}}\right)^{1/3}R$, where $\rho_{i}=0$, acts as the boundary of this region. Within the green portion of the diagram we find the maximum compactness bound~\eqref{Eq:CompLimRhoP}. Once we move below said curve, thus allowing for solutions with $\rho_{i}<0$, it is always possible to find broad regions (shaded in blue) of perfectly regular stars, bounded by $R_{+}$ and $R_{-}$. The maximum compactness of these specific stars can be \textit{arbitrarily close} to the black hole compactness. 

\begin{figure}
    \centering
    \includegraphics[width=0.75\linewidth]{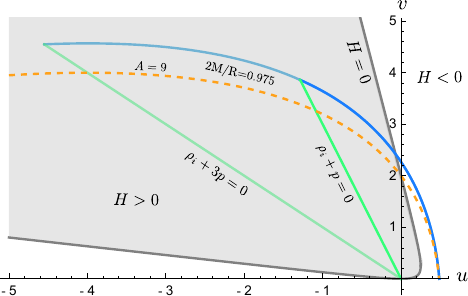}
    \caption{$(u,v)$ diagram of bilayered stars with negative energy densities. 
    The $v(u)$ curves from the inner layer appear in green and those from the outer layer in blue. We have plotted two example solutions satisfying $\rho_{i}+p=0$ and $\rho_{i}+3p=0$ and which surpass Bondi's limit. The compactness of these solutions can be arbitrarily close to the black hole limit.}
    \label{Fig:UVNeg}
\end{figure}

These solutions have a complementary interpretation in terms of the $(u,v)$ variables used in Subsection~\ref{SubSec:Bondi_Limit}. The $v(u)$ curve describing the constant-density core solutions is given by
\begin{equation}\label{Eq:vuNegrho}
    v=-\frac{u\left(v_{0}+3\sqrt{3-6u}\right)}{v_{0}+\sqrt{3-6u}},
\end{equation}
where $v_{0}$ is an integration constant related to the central pressure via
\begin{equation}
    v_{0}=-\frac{3\sqrt{3}\left(\rho_{i}+p(0)\right)}{\rho_{i}+3p(0)}
\end{equation}
The constant-pressure curves $\rho+p_{i}=0$ and $\rho+3p_{i}=0$ are just the straight lines $v=-3u$ and $v=-u$ in the $(u,v)$ diagram from Fig.~\ref{Fig:uvDiagram}. These are obtained by evaluating~\eqref{Eq:vuNegrho} in the limits $v_{0}\to0$ and $v_{0}\to\pm\infty$, respectively. Since these $v$ lines always intersect the $P_{A}$ parabolas~\eqref{Eq:PACurves} at some $u<0$ for any $A>1$, it is possible to match them to an outer constant-density layer so that the compactness approaches the black hole compactness $u=1/2$ $(A\to\infty)$ as much as desired. Fig. \ref{Fig:UVNeg} shows the $v(u)$ curves for the two models with constant-pressure inner layers that we presented above.

\section{AdS Stars and Einstein Static stars} 
\label{Sec:AdSStars}
Within the bilayered models with negative energy density at the core, we found two with rather special characteristics whose pressure profiles were plotted in Figs.~\ref{Fig:RP}--\ref{Fig:RM}. Inspired by these solutions, and considering stars with a compactness close to but below the black hole limit ($8/9<2M/R\sim 1$), we identify configurations with thin external crusts that fall into one of the following two categories: 
\begin{enumerate}
\item In the core they have a negative density
that makes the pressure exactly constant, such that \mbox{$p(r)=p_i = -\rho_i$}. In the limit in which the outer thick layer becomes infinitesimally thin (distributional), we shall call these stellar configurations \textit{AdS stars}. See Fig.~\ref{Fig:RP} for a particular example in which the outer shell is thick.
\item  In the core they have a negative density that makes the pressure exactly constant, such that \mbox{$p(r)=p_i=-\rho_i/3$}.
In the limit in which the outer thick layer becomes infinitesimally thin (distributional), we shall call these stellar configurations \textit{Einstein static stars}. See Fig.~\ref{Fig:RM} for a particular example in which the outer shell is thick.
\end{enumerate}
AdS stars are constituted by an anti-de Sitter interior glued through a thin shell to a Schwarzschild exterior. The geometry can be built through a cut and paste procedure with the help of Israel junction conditions. The interesting thing here is that the qualitative behavior of this \emph{ad hoc} geometry constructed through this surgery procedure appears naturally within the set of exact regular bilayered models, the only difference being that in that case the shell-like behavior is absent and the transition between the inside core and the outside is smooth.

\begin{figure}[H]
\begin{center}
\includegraphics[width=0.35 \textwidth, height=0.33 \textwidth]{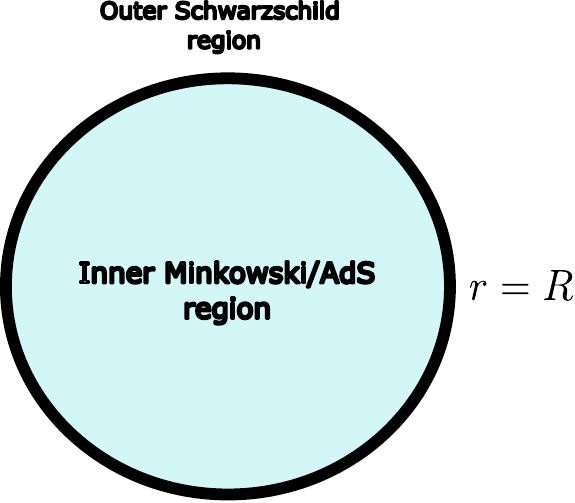}
\caption{Pictorial representation of the matching of the inner Minkowskian or AdS core with the external Schwarzschild region through a spherically symmetric thin shell.}
\label{Fig:Shell}
\end{center}
\end{figure}

An additional feature that AdS and Einstein stars display is that the (distributional) matter supporting the configuration is a non-perfect fluid, e.g., a fluid exhibiting anisotropic pressures. The stress-energy tensor can be expressed in general as:
\begin{align}
    T_{ab} =  \rho T_{a} T_{b} + p_r R_a R_b + p_t \left( \Theta_a \Theta_b + \Phi_a \Phi_b \right),
    \label{Eq:GenericFluid}
\end{align}
where $\{ T_a, R_a, \Theta_a, \Phi_b \}$ represents a tetrad adapted to the symmetry of the problem, and we have introduced the density $\rho$, the radial pressure $p_r$, and the tangential pressure $p_t$ (notice that the two tangential pressures need to be equal in order for the spherical symmetry to be preserved). The setup that we are considering is depicted in Fig.~\ref{Fig:Shell}.
Let us specify the construction in detail: Take a spherical shell with radius $R$, such that its proper area is $4 \pi R^2$. Let the spacetime outside the shell be Schwarzschild with a mass $M$, and AdS inside the shell. We denote the Schwarzschild (outside) patch with a $+$ sign, and the AdS (inside) patch with a $-$ sign. The metric is now given by the two line elements in Eq.~\eqref{Eq:thinshellbulks}) with $f_{+}(r_{+}) = 1 - 2M/r_{+}$, $h_{+}(r_{+})=1/f_{+}(r_{+})$ and $f_{-}(r_{-}) = 1 + \abs{\lambda} r^2$, $h_{-}(r_{-})=1/f_{-}(r_{-})$. The real parameter $\lambda$ corresponds to a negative cosmological constant. We locate the shell at $r_{+} = r_{-} = R$. The matching is worked out in detail in Appendix~\ref{App:AnisotropicShell}, and we find that the shell displays a distributional stress-energy tensor with the following densities and pressures:
\begin{align}
    & \rho = \sigma \delta (r - R), \\
    & p_r = 0, \\
    & p_t = \tilde{p}_t \delta(r -R),
\end{align}
with $\sigma$ and $\tilde{p}_t$ given by
\begin{align}
     \sigma & = \frac{1}{4 \pi R} \left(\sqrt{1+|\lambda|R^2} - \sqrt{1-\frac{2M}{R}} \right), \nonumber \\
     \tilde{p}_t & = -\frac{1}{8  \pi R } \frac{d}{dR} \left(
     R\sqrt{1+|\lambda|R^2} - R \sqrt{1-\frac{2M}{R}} \right).
    \label{Eq:Densities}
\end{align}
In the crust of the bilayered model the pressure increases from zero to $p_i$. In the distributional limit the pressure exhibits just a jump. The closer the compactness is to 1, the larger this jump in pressure becomes.  In this work, we have introduced  bilayered models to illustrate their potentiality to generate configurations surpassing the standard Buchdahl limit. However, it is remarkable that the structure of AdS stars is precisely the 
backbone of the semiclassical stellar solutions found previously in~\cite{Arrecheaetal2021,Arrecheaetal2023}. The only energy-momentum tensor supporting the stellar configurations reported there is that of a perfect fluid of constant density, together with the semiclassical renormalized stress-energy tensor (RSET) that they generate.  

In this self-consistent semiclassical solutions there is always a classical positive energy density which in the core is overcome by a negative semiclassical contribution from the RSET. In the limit in which the crust is very thin, these stars can be approximated by AdS stars. In that sense, one can consider AdS stars as idealized models of semiclassical ultracompact stars whose energy momentum tensor is the sum of a perfect fluid and the RSET of a quantum scalar field.
It is also interesting to point out that these AdS stars can be interpreted as an inverted version of the well-known gravastar model of Mazur and Mottola~\cite{MazurMottola2023}, in the sense that gravastars have a de Sitter interior instead of an AdS one.

We also have to mention here the similarities and differences between our AdS stars and the AdS black shells of~\cite{Danielssonetal2017}. The biggest difference comes from the compactness. In the AdS black shell construction, the shell is located at around the Buchdahl radius $R=9M/4$. In our case, the shell is located as near to the black hole limit as desired. 

Einstein stars are those in which the core has constant positive pressure satisfying \mbox{$\rho_i = -3 p_i$}. We have called them Einstein stars because of their similarities with Einstein static cosmological model~\cite{Einstein1917}.
In that model, a 3-sphere filled with a pressureless dust, representing the matter content of the universe, is counterbalanced by the presence of a cosmological constant. As its cosmological acceleration is zero, we have that $\rho_{\rm total}+ 3p_{\rm total}=0$. In the cosmological case we have that $\rho_{\rm dust}+ \rho_\Lambda +3p_\Lambda = \rho_{\rm dust} - \Lambda/(4 \pi)=0$, with $\Lambda$ the cosmological constant. In the case of our semiclassical stars we have 
\begin{equation}
    \rho_{\rm class} + \rho_{\rm semiclass} + 3 p_{\rm semiclass}=0,
\end{equation}
with $\rho_{\rm semiclass}$ being the only negative term. Geometrically speaking, these stars have a constant non-zero redshift core.

These models could serve as simple templates of black hole mimickers inspired by semiclassical effects. In fact, more sophisticated models can be constructed with a nesting of two of these toy models, as done in~\cite{Jampolski2023} with gravastars. One central element responsible for both the features of their ringdown~\cite{CardosoPani2019} and their associated shadows~\cite{Ayzenbergetal2023} are the redshift and time delay effects suffered by light rays crossing their interiors. While the AdS star has a redshift function that decreases towards the centre, the one associated to the Einstein static star is fully constant. This would result in radically different crossing times for null rays and, in turn, to very different observational signatures both in the electromagnetic and gravitational-wave spectra.

\section{Anisotropy} 
\label{Sec:Anisotropy}

For the sake of completeness in our analysis, in this section we will discuss the isolated effect of anisotropy in bypassing the Buchdahl limit. To this end,
it is convenient to focus on a situation in which there are no violations of energy conditions arising from an interior core. For that purpose, we can analyze a toy model of a shell matching an exterior Schwarzschild spacetime with mass $M$ to an interior flat spacetime (i.e., Schwarzschild with $M =0$). The setup is the same as the one depicted in Fig.~\ref{Fig:Shell}, but considering the core to be empty instead of AdS. This simply corresponds to setting the AdS constant $\lambda = 0$ from previous results, and we find the following energy and tangential pressure densities:
\begin{align}
     \sigma = \frac{1}{4 \pi R} \left( 1 - \sqrt{f_{+}(R)} \right), \quad
     \tilde{p}_t = \frac{1}{16  \pi R \sqrt{f_{+}(R)}} \left(- 2\sqrt{f_{+}(R)} + 2 f_{+}(R) + f_{+}'(R) R  \right).
    \label{Eq:Densities2}
\end{align}
Given that $\tilde{p}_t $ blows up at $R= 2M$, and that it goes to zero faster than $\sigma$ as $R \to \infty$, there is a crossover at a given $R = R_0$ for which $\sigma = \tilde{p}_t$. This means that we can expect violations of energy conditions. In fact, the DEC requires
\begin{align}
    \rho \geq p_t. 
\end{align}
Given that the two of them have the same distributional behavior, for our particular example this would imply
\begin{align}
    \sigma \geq \tilde{p}_t.
\end{align}
If we plug in the explicit values of $\sigma$ and $\tilde p_t$ given in Eq.~\eqref{Eq:Densities}, we find the bound:

\begin{align}
    C= \frac{2M}{R} \leq \frac{24}{25} = 0.96.
\end{align}
This bound is less stringent than the Buchdahl bound for isotropic situations, $C<8/9 \sim 0.888$ ~\cite{Buchdahl1959}. But it is more stringent than the bound that follows from just demanding the positivity of the $g_{rr}$ component of the metric---the black hole limit $C<1$. 

In~\cite{Ivanov2002}, bounds were derived for the maximum redshift that the surfaces of anisotropic stars can have under several assumptions, namely:
\begin{align}
    s = \frac{1}{(1-2M/R)^{1/2}}-1 \leq s_{\text{max}},
\end{align}
where the value of the $s_{\text{max}}$ depends on the energy condition that is chosen. In terms of the compactness, we have that the bounds are of the form:
\begin{align}
    \frac{2M}{R} \leq \frac{s_{\text{max}}(s_{\text{max}}+2)}{(1+s_{\text{max}})^2}.
\end{align}
In~\cite{Ivanov2002}, the bounds are derived assuming that the radial pressure is strictly positive (for stability reasons). It is also assumed that $\rho$ is positive and a decreasing function (although it can stay constant for a given interval). Assuming the DEC is obeyed, the upper bound for the redshift found is:
\begin{align}
    s_{\text{max}}^{\text{DEC}} = 5.421 \rightarrow \frac{2M}{R} \leq 0.974,
\end{align}
whereas imposing the condition that $\rho \geq 2 p_t$ leads to a more stringent bound\footnote{We believe that although Ivanov refers to this condition as the Strong Energy Condition, this is actually incorrect. If the condition $p_t \geq p_r$ is satisfied and the density and pressures are positive as they discuss, the SEC is actually much more restrictive.}:
\begin{align}
    s_{\text{max}}^{\text{SEC}} = 3.842 \rightarrow \frac{2M}{R} \leq 0.958.
\end{align}
We can compare these bounds with the shell model that we are considering, and we realize that the shell model begins to violate the DEC whenever the compactness is greater than $2M/R = 0.96$, which still obeys the bound $2M/R \leq 0.974$ found by Ivanov~\cite{Ivanov2002}. Further results that put bounds on the compactness of the objects with similar assumptions are the results reported in~\cite{Barraco2003,Boehmer2006,Andreasson2007,Urbano2018}, see also~\cite{Rosa2020}. Similar analyses for elastic self-gravitating bodies were carried on in~\cite{Alho2021,Alho2022,Alho2023,Alho2023b}, reaching similar conclusions. In~\cite{Barraco2003}, under roughly speaking the same assumptions as Ivanov's, it is shown that anisotropic models fall into two classes according to whether $p_r > p_t$ or $p_t >p _r$. In the case $p_r > p_t$, they manage to show that the $p_r$ function is strictly greater than the one in a fiduciary isotropic model with the same function $m(r)$. This means that these stars cannot be super-Buchdahl; actually, isotropic stars obeying DEC need to be less compact than Buchdahl~\cite{Barraco2002}. Regarding the case $p_r < p_t$, they show that the radial pressure is always bigger than the one in the fiduciary isotropic model with the same Misner-Sharp mass profile. Imposing the DEC, they derive bounds that are consistent with Ivanov's. In~\cite{Boehmer2006}, under some additional assumptions (about surface density behavior motivated on physical grounds), slightly sharper bounds are proved. Finally, in~\cite{Lindblom1984,Urbano2018} it is shown that imposing a causality condition, namely that the speed of sound is smaller than the speed of light, there is a maximum compactness that can be reached, which turns out to be below Buchdahl's limit, namely $2M/R \lesssim 0.768$. 

In any case, the simple example exhibited in this section illustrates that in the absence of energy conditions, anisotropic configurations can be made as compact as desired. Energy conditions set a limit to the compactness, which is nevertheless far from Buchdahl's, leaving the door open to having extremely compact objects that are close to forming event horizons, and are even allowed to display photon rings (since that only requires \mbox{$2M/R > 2/3 \sim 0.666$}).

\section{Discussion and future work}
\label{Sec:Discussion}

In this article, we have revisited the different ways in which Buchdahl's limit can be bypassed, relaxing separately each of the main assumptions invoked in the theorem: namely, the outward-decreasing monotonicity of the density profile, and the isotropy of the stress-energy tensor. Regarding the former, we reviewed the bound obtained by Bondi~\cite{Bondi1964}, which was later rigorously proven by Karageorgis and Stalker~\cite{Karageorgis2007}, showing 
that for non-decreasing (yet everywhere non-negative) density profiles, a compactness limit different from Buchdahl's arises. 

Here, we have found geometrically smooth toy models that allow us to approach this limit as closely as desired. We also show that as long as we allow negative densities (and hence violations of the weak energy condition), our toy model can describe objects arbitrarily close to the compactness of a black hole, in agreement with Bondi’s results. Among the possibilities explored, we derived a particularly simple configuration consisting of an anti-de Sitter interior and a thin shell matched to an exterior Schwarzschild metric, which we have called \textit{AdS star}, and which bears a close resemblance to solutions obtained in semiclassical gravity~\cite{Arrecheaetal2023}. On the other hand, we also introduced a very simple toy model of an anisotropic shell that allowed us to show that, without further restrictions (such as, e.g., energy conditions), one can conceive objects with compactness arbitrarily close to the black hole limit. The imposition of some energy condition still allows the surpassing of Buchdahl's limit, but places new upper bounds on the compactness.

Once we allow violations of energy conditions, we need to be careful, since we are opening the door to a plethora of pathological behaviors. For instance, violations of the dominant energy condition are often interpreted as leading to acausal behavior. The DEC can be understood as representing that the speed of the energy flow of the matter field is always subluminal. This is manifested in a theorem by Hawking~\cite{Hawking1970,Hawking1973}  which ensures that if the energy-momentum tensor obeys the DEC and it vanishes on a piece of a Cauchy slice, then it vanishes also on the domain of influence of such piece of the Cauchy slice. This can be interpreted as saying that signals obeying the DEC cannot propagate faster than light~\cite{Wong2010}. However, that the shell (or the constant-density star) itself does not obey the DEC does not necessarily mean that its perturbations are necessarily acausal, since the perturbations with respect to a configuration that violates the DEC are not guaranteed to violate the DEC as well. To put it explicitly, given our configuration with background values of the density and pressure $\bar{\rho}, \bar{p}$ that violate the DEC (i.e., $\abs{\bar{\rho}} \leq \abs{\bar{p}}$), the perturbations $\delta \rho, \delta p$ do not necessarily obey $ \abs{\delta p} \leq \abs{\delta \rho}$. Although it might be useful to understand this in terms of a background and a perturbation energy-momentum tensor, such splitting comes with an intrinsic arbitrariness that makes it hard to be precise about in what sense background configurations may violate the DEC without one entering into pathological behavior for perturbations respecting it. Furthermore, the superluminal behavior that is tied to violations of the DEC may not necessarily be problematic, as there are examples of situations in which apparently superluminal behaviors do not necessarily entail a pathology~\cite{Babichev2007,Geroch2010,Barcelo2022}. We believe that the DEC violation in this case would point to an instability of the background (which in the shell case that we analyze is natural since the object would tend to collapse). The theorems in which the DEC is used, namely, to prove many of the results concerning no-hair theorems, black hole mechanics, and to ensure the well-posedness of the initial value problem in its general formulation~\cite{Curiel2014}, do not seem to contradict this interpretation.

When we allow non-monotonically decreasing density profiles, we may expect that if the matter content is that of an ordinary fluid, it will display an unstable behavior, but no energy conditions need to be violated, \textit{a priori}. If we go further and allow negative energy densities, causally pathological behavior might arise due to the violation of the WEC. However, if such behavior of the effective densities and pressures is motivated by an underlying healthy physical mechanism, e.g. semiclassical physics, the pathological behavior might not arise. In fact, such behavior is expected from assuming that perturbations with respect to the background constrain the density and pressures in a similar way the background ones do, without this being necessarily the case. For example, a horizonless alternative to GR black holes has recently appeared within semiclassical theories of gravity, where vacuum polarization effects in stars approaching Buchdahl's limit predict the appearance of outwards increasing (and even negative) energy densities, allowing to bypass Buchdahl's theorem~\cite{Arrecheaetal2023}. The bilayered stars proposed in this work capture the core properties of these semiclassical solutions while being devoid of the intricacies of semiclassical analyses, which might foster their phenomenological scrutiny in the future.

\begin{acknowledgements}
The authors thank  Jos\'e M.M. Senovilla, Luis J. Garay, Ignacio Reyes and Ana Alonso-Serrano for helpful discussions. 
We also thanks Shahar Hod and H\aa kan Andr\'easson for informing us about important references. Financial support was provided by the Spanish Government through the Grants No. PID2020-118159GB-C43 and PID2023-149018NB-C43 (funded by MCIN/AEI/10.13039/501100011033), and by the Junta de Andaluc\'{\i}a through the project FQM219. GGM is funded by the Spanish Government fellowship FPU20/01684. CB and GGM acknowledge financial support from the Severo Ochoa grant CEX2021-001131-S funded by MCIN/AEI/ 10.13039/501100011033. JPG acknowledges the support of a Mike and Ophelia Lazaridis Fellowship, as well as the support of a fellowship from ``La Caixa'' Foundation (ID 100010434, code LCF/BQ/AA20/11820043). 
\end{acknowledgements}

\newpage

\appendix 

\section{Fluid spheres in GR}
\label{App:FluidSpheres}

Let us consider the line element in Eq.~\eqref{Eq:LineElement2}, and the solution corresponding to a perfect fluid with constant density profile, namely:
\begin{align}
    \rho(r) = \begin{cases}
      \rho_0  & r \leq R\\
      0 & r>R
    \end{cases}. 
\end{align}
For such density profile, the mass function $m(r)$ takes the form
\begin{align}
    m(r) = \begin{cases}
      \frac{4}{3} \pi r^3 \rho_0  & r \leq R\\
      \frac{4}{3} \pi R^3 \rho_0 & r>R
    \end{cases},
\end{align}
as a straightforward integration demonstrates, and we have a relation among the parameters $\left( M,R,\rho_0 \right)$ which is precisely $M = (4/3) \pi R^3 \rho_0$. The pressure and the redshift function also admit explicit expressions given by
\begin{align}
    P(r) = \rho_0 \left[ \frac{(1-2M/R)^{1/2} - (1-2Mr^2/R^3)^{1/2}}{(1-2Mr^2/R^3)^{1/2} - 3 ( 1 - 2M/R)^{1/2} }\right]
    \label{Eq:ConstantDens_Pressure}
\end{align}
and
\begin{align}
    \Phi(r) = \log \left[ \frac{3}{2} \left( 1 - 2 M /R \right)^{1/2} - \frac{1}{2} \left( 1 - 2 Mr^2/R^3\right)^{1/2} \right].
    \label{Eq:ConstantDens_Redshift}
\end{align}
for $r\leq R$. Everywhere else, the pressure is zero, and the redshift function matches its Schwarzschild counterpart, given in Eq.~\eqref{Eq:SchwMetric}. We can particularize the pressure from Eq.~\eqref{Eq:ConstantDens_Pressure} to $r = 0$ to realize that the value of the pressure at the core is given by
\begin{align}
    P_c = P (0) = \rho_0 \left[ \frac{1 - (1-2M/R)^{1/2}}{3 (1-2M/R)^{1/2} - 1 } \right],
\end{align}
which clearly blows up when $R \rightarrow 9 M/4$, i.e., as Buchdahl's limit is approached. Furthermore, it is convenient to rewrite the maximum mass allowed for a star of a given density in terms of the density instead of the radius, namely:
\begin{align}
    M_{\text{Buchdahl}} = \frac{4}{9} R = \frac{4}{9 (3 \pi)^{1/2}} \rho_0^{-1/2}. 
\end{align} 
Even though the density is not a continuous function because of the jump that it displays at $ r =R$, $h$ is still continuous, which is what is required for the proof of Buchdahl's theorem. In fact, although in this case we are considering a discontinuous function, it is always possible to understand this density profile as a limit that can be approached within a parametric family of smooth $C^{\infty}$ functions $\rho_{\ell} (r)$, such as
\begin{align}
    \rho_{\ell} (r) =  \begin{cases}
      \rho_0\, e^{ \frac{4 \ell^2}{R^2} + \frac{\ell^2}{r (r-R)}  } &  r < R \\
      0 & r > R 
    \end{cases},  
    \label{Eq:RegDensity}
\end{align}
which in the limit $\ell \rightarrow 0^+ $ reduces to a Heaviside function with density $\rho_0$ in the interior $r\in (0, R)$  and $0$ in the external region. 

\section{Fully anisotropic shell matching spherically symmetric spacetimes}
\label{App:AnisotropicShell}

Let us consider a static and spherically symmetric spacetime that is foliated in spheres of proper radius $r$. Consider a line element of the form:
\begin{align}
    ds^2 = -f(r) dt^2 + h(r) dr^2 +r^2 d \Omega_2.
\end{align}
We can consider that the function $f(r)$ and $h(r)$ are piecewise defined, namely that we have:
\begin{align}
    f(r) = \begin{cases}
      f_{-} (r)  & r < R \\
      f_{+} (r)  & r > R 
    \end{cases}, \qquad
        h(r) = \begin{cases}
      h_{-} (r)  & r < R \\
      h_{+} (r)  & r > R 
    \end{cases},
\end{align}

The first of Israel junction conditions states that the metric needs to be continuous and hence the induced metric on the surface cannot have a discontinuity. The induced metric for the surface (whose components we denote as $h_{ab}$) can be computed from the outside and inside metrics to obtain:
\begin{align}
    & ds^2_{+} = - f_{+}(R) \left(\frac{dt_{+}}{d \tau} \right)^2 d\tau^2 + R^2 d \Omega_{+,2} ^2, \\
    & ds^2_{-} = - f_{-}(R) \left(\frac{dt_{-}}{d \tau} \right)^2 d\tau^2 + R^2 d \Omega_{-,2} ^2,
\end{align}
where $\tau$ is the proper time of constant $\Theta_{-}$ observers. Continuity of the metric ensures that we can take an orthonormal basis on the surface such that $\Theta^{a} \partial_a= \partial_{\theta^{+}} = \partial_{\theta^{-}}$, \mbox{$\Phi^a \partial_a= \partial_{\varphi^{+}} = \partial_{\varphi^{-}}$}, and 
\begin{align}
    f_{+}(R) \left( \frac{d t_{+}}{d \tau} \right)^2 = f_{-} (R) \left( \frac{d t_{-}}{d \tau} \right)^2,
\end{align}
which translates into the condition $\sqrt{f_{+}(R)} t_{+} = \sqrt{f_{-} (R)} t_{-}$, choosing the same origin for the time coordinate. In other words, the induced metric is given by
\begin{align}
    ds^2 = - d \tau^2 + R^2 d \Omega^2,
\end{align}
since we can identify the angular coordinates and the proper time long the shell is adapted to the redshift functions from the inside and the outside. The vector tangent to the surface $u = \partial_{\tau}$ can be expressed as:
\begin{align}
    & u = \partial_{\tau}  = \frac{\partial t_{-}}{\partial \tau} \partial_{t_{-}} = \frac{1}{\sqrt{{f_{-} (R)}}} \partial_{t_{-}}, \\
    & u = \partial_{\tau} = \frac{\partial t_{+}}{\partial \tau} \partial_{t_{+}} = \frac{1}{\sqrt{{f_{+} (R)}}} \partial_{t_{+}}.
\end{align}
We now need $n$, the unit normal vector to the surface which is orthogonal to $u$. It is given by
\begin{align}
    & n^{+}_{a} = \sqrt{h_{+}(r)} \delta^{r}_{\ a} ,
    & n^{-}_{a} = \sqrt{h_{-}(r)}\delta^{r}_{\ a}. 
\end{align}
To compute the extrinsic curvature, we need $\nabla_{b} n_a$, and we can extract the different components of the extrinsic curvature through
\begin{align}
    & K_{\tau \tau} = \nabla_b n_a u^b u^a, \\ 
    & K_{\theta \theta} = \nabla_b n_a \Theta^a \Theta^b, \\
    & K_{\Phi \Phi} = \nabla_b n_a \Phi^a \Phi^b. 
\end{align}
Computed from the outside the extrinsic curvature gives:
\begin{align}
    & K^{+}_{\tau \tau} = - \frac{1}{2} \frac{f'_{+}(R)}{f_{+}(R) \sqrt{h_{+}(R)}}, \\ 
    & K^{+}_{\theta \theta} =   \frac{R}{\sqrt{h_{+}(R)}}, \\
    & K^{+}_{\Phi \Phi} =  \frac{R \sin^2 \theta }{\sqrt{h_{+}(R)}} , \\
    & K^{+} = K^{+}_{ab}h^{ab} = \frac{1}{2} \frac{f'_{+}(R)}{f_{+}(R) \sqrt{h_{+}(R)}} + 2 \frac{R}{\sqrt{h_{+}(R)}}, 
\end{align}
whereas from the inside we simply need to replace $f_{+}$ and $h_{+}$ by $f_{-}$ and $h_{-}$ respectively:
\begin{align}
    & K^{-}_{\tau \tau} = - \frac{1}{2} \frac{f'_{-}(R)}{f_{-}(R) \sqrt{h_{-}(R)}}, \\ 
    & K^{-}_{\theta \theta} =   \frac{R}{\sqrt{h_{-}(R)}}, \\
    & K^{-}_{\Phi \Phi} =  \frac{R \sin^2 \theta }{\sqrt{h_{-}(R)}} , \\
    & K^{-} = K^{-}_{ab}h^{ab} = \frac{1}{2} \frac{f'_{-}(R)}{f_{-}(R) \sqrt{h_{-}(R)}} + 2 \frac{R}{\sqrt{h_{-}(R)}}, 
\end{align}
The second junction condition dictates that the jump in the extrinsic curvature is directly proportional to the distributional energy-momentum tensor. Explicitly, we have:
\begin{align}
    8  \pi S^{a}_{\ b} = - \left( \left[ \left[ K^{a}_{\ b} \right] \right] -  \left[ \left[ K\right] \right] \delta^a_{\ b} \right),
\end{align}
where we have introduced the notation $\left[ \left[ \mathcal{O} \right] \right] = \mathcal{O}^{+} - \mathcal{O}^{-}$, representing precisely the jump in the extrinsic curvatures. We can compute it to find:
\begin{align}
    & \left[ \left[ K^{\tau}_{\ \tau}\right] \right] =  \frac{1}{2} \frac{f'_{+}(R)}{f_{+}(R) \sqrt{h_{+}(R)}} -  \frac{1}{2} \frac{f'_{-}(R)}{f_{-}(R) \sqrt{h_{-}(R)}}, \\ 
    & \left[ \left[ K^{\theta}_{\ \theta}\right] \right] =  \frac{1}{R \sqrt{h_{+}(R)}} - \frac{1}{R \sqrt{h_{-}(R)}} , \\
    & \left[ \left[ K^{\Phi}_{\ \Phi}\right] \right] =  \frac{1}{R \sqrt{h_{+}(R)}} - \frac{1}{R \sqrt{h_{-}(R)}}, \\
    & \left[ \left[ K \right] \right] = \frac{1}{2} \left( \frac{f'_{+}(R)}{f_{+}(R) \sqrt{h_{+}(R)}} - \frac{f'_{-}(R)}{f_{-}(R) \sqrt{h_{-}(R)}} \right) + \frac{2}{R} \left[ \frac{1}{ \sqrt{h_{+} (R)}} - \frac{1}{\sqrt{ h_{-} (R)}} \right]. 
\end{align}
From this expression we can determine $S^{a}_{\ b}$. For our purpose, it is interesting that we can express the tensor as that of a perfect fluid,
\begin{align}
    S_{ab} = \sigma u_a u_b + \tilde{p}_t \left( h_{ab} + u_a u_b \right),
\end{align}
where $\sigma$ and $\tilde{p}_t$ represent the surface energy density and the surface tangential pressure. We have $S^{\tau}_{\ \tau} = - \sigma$ and $S^{\theta}_{\ \theta} = \tilde{p}_{t}$. They are given by
\begin{align}
     \sigma & = \frac{1}{4 \pi R} \left( \frac{1}{\sqrt{h_{-}(R)}} - \frac{1}{\sqrt{h_{+}(R)}} \right), \\
     \tilde{p}_t & = \frac{1}{8 \pi R} \left[ \frac{1}{ \sqrt{h_{+} (R)}} - \frac{1}{\sqrt{ h_{-} (R)}} \right] + \frac{1}{16 \pi} \left( \frac{f'_{+}(R)}{f_{+}(R) \sqrt{h_{+}(R)}} - \frac{f'_{-}(R)}{f_{-}(R) \sqrt{h_{-}(R)}} \right).
    \label{Eq:DensitiesApp}
\end{align}

In general, whenever $\tilde{p}_t \neq 0$, the presence of a thin-shell automatically introduces anisotropic pressures. However, the special situations in which $\tilde{p}_t =0$ do not introduce any anisotropic pressure into play, since the only component of the energy-momentum tensor that does not identically vanish corresponds to the energy density, which exhibits a distributional behavior.

\bibliography{buchdahl_biblio}

\begin{thebibliography}{52}%
\makeatletter
\providecommand \@ifxundefined [1]{%
 \@ifx{#1\undefined}
}%
\providecommand \@ifnum [1]{%
 \ifnum #1\expandafter \@firstoftwo
 \else \expandafter \@secondoftwo
 \fi
}%
\providecommand \@ifx [1]{%
 \ifx #1\expandafter \@firstoftwo
 \else \expandafter \@secondoftwo
 \fi
}%
\providecommand \natexlab [1]{#1}%
\providecommand \enquote  [1]{``#1''}%
\providecommand \bibnamefont  [1]{#1}%
\providecommand \bibfnamefont [1]{#1}%
\providecommand \citenamefont [1]{#1}%
\providecommand \href@noop [0]{\@secondoftwo}%
\providecommand \href [0]{\begingroup \@sanitize@url \@href}%
\providecommand \@href[1]{\@@startlink{#1}\@@href}%
\providecommand \@@href[1]{\endgroup#1\@@endlink}%
\providecommand \@sanitize@url [0]{\catcode `\\12\catcode `\$12\catcode
  `\&12\catcode `\#12\catcode `\^12\catcode `\_12\catcode `\%12\relax}%
\providecommand \@@startlink[1]{}%
\providecommand \@@endlink[0]{}%
\providecommand \url  [0]{\begingroup\@sanitize@url \@url }%
\providecommand \@url [1]{\endgroup\@href {#1}{\urlprefix }}%
\providecommand \urlprefix  [0]{URL }%
\providecommand \Eprint [0]{\href }%
\providecommand \doibase [0]{http://dx.doi.org/}%
\providecommand \selectlanguage [0]{\@gobble}%
\providecommand \bibinfo  [0]{\@secondoftwo}%
\providecommand \bibfield  [0]{\@secondoftwo}%
\providecommand \translation [1]{[#1]}%
\providecommand \BibitemOpen [0]{}%
\providecommand \bibitemStop [0]{}%
\providecommand \bibitemNoStop [0]{.\EOS\space}%
\providecommand \EOS [0]{\spacefactor3000\relax}%
\providecommand \BibitemShut  [1]{\csname bibitem#1\endcsname}%
\let\auto@bib@innerbib\@empty
\bibitem [{\citenamefont {Collmar}\ \emph {et~al.}(1998)\citenamefont
  {Collmar}, \citenamefont {Straumann}, \citenamefont {Chakrabarti},
  \citenamefont {'t~Hooft}, \citenamefont {Seidel},\ and\ \citenamefont
  {Israel}}]{Werneretal1998}%
  \BibitemOpen
  \bibfield  {author} {\bibinfo {author} {\bibfnamefont {W.}~\bibnamefont
  {Collmar}}, \bibinfo {author} {\bibfnamefont {N.}~\bibnamefont {Straumann}},
  \bibinfo {author} {\bibfnamefont {S.~K.}\ \bibnamefont {Chakrabarti}},
  \bibinfo {author} {\bibfnamefont {G.}~\bibnamefont {'t~Hooft}}, \bibinfo
  {author} {\bibfnamefont {E.}~\bibnamefont {Seidel}}, \ and\ \bibinfo {author}
  {\bibfnamefont {W.}~\bibnamefont {Israel}},\ }in\ \href@noop {} {\emph
  {\bibinfo {booktitle} {Black Holes: Theory and Observation}}},\ \bibinfo
  {editor} {edited by\ \bibinfo {editor} {\bibfnamefont {F.~W.}\ \bibnamefont
  {Hehl}}, \bibinfo {editor} {\bibfnamefont {C.}~\bibnamefont {Kiefer}}, \ and\
  \bibinfo {editor} {\bibfnamefont {R.~J.}\ \bibnamefont {Metzler}}}\ (\bibinfo
   {publisher} {Springer Berlin Heidelberg},\ \bibinfo {address} {Berlin,
  Heidelberg},\ \bibinfo {year} {1998})\ pp.\ \bibinfo {pages}
  {481--489}\BibitemShut {NoStop}%
\bibitem [{\citenamefont {Abramowicz}\ \emph {et~al.}(2002)\citenamefont
  {Abramowicz}, \citenamefont {Kluzniak},\ and\ \citenamefont
  {Lasota}}]{Abramowicz2002}%
  \BibitemOpen
  \bibfield  {author} {\bibinfo {author} {\bibfnamefont {M.~A.}\ \bibnamefont
  {Abramowicz}}, \bibinfo {author} {\bibfnamefont {W.}~\bibnamefont
  {Kluzniak}}, \ and\ \bibinfo {author} {\bibfnamefont {J.-P.}\ \bibnamefont
  {Lasota}},\ }\href {\doibase 10.1051/0004-6361:20021645} {\bibfield
  {journal} {\bibinfo  {journal} {Astron. Astrophys.}\ }\textbf {\bibinfo
  {volume} {396}},\ \bibinfo {pages} {L31} (\bibinfo {year} {2002})},\ \Eprint
  {http://arxiv.org/abs/astro-ph/0207270} {arXiv:astro-ph/0207270} \BibitemShut
  {NoStop}%
\bibitem [{\citenamefont {Carballo-Rubio}\ \emph {et~al.}(2022)\citenamefont
  {Carballo-Rubio}, \citenamefont {Cardoso},\ and\ \citenamefont
  {Younsi}}]{Carballo-Rubio2022}%
  \BibitemOpen
  \bibfield  {author} {\bibinfo {author} {\bibfnamefont {R.}~\bibnamefont
  {Carballo-Rubio}}, \bibinfo {author} {\bibfnamefont {V.}~\bibnamefont
  {Cardoso}}, \ and\ \bibinfo {author} {\bibfnamefont {Z.}~\bibnamefont
  {Younsi}},\ }\href {\doibase 10.1103/PhysRevD.106.084038} {\bibfield
  {journal} {\bibinfo  {journal} {Phys. Rev. D}\ }\textbf {\bibinfo {volume}
  {106}},\ \bibinfo {pages} {084038} (\bibinfo {year} {2022})},\ \Eprint
  {http://arxiv.org/abs/2208.00704} {arXiv:2208.00704 [gr-qc]} \BibitemShut
  {NoStop}%
\bibitem [{\citenamefont {{Nauenberg}}(2008)}]{Nauenberg2008}%
  \BibitemOpen
  \bibfield  {author} {\bibinfo {author} {\bibfnamefont {M.}~\bibnamefont
  {{Nauenberg}}},\ }\href {\doibase 10.1177/002182860803900302} {\bibfield
  {journal} {\bibinfo  {journal} {Journal for the History of Astronomy}\
  }\textbf {\bibinfo {volume} {39}},\ \bibinfo {pages} {297} (\bibinfo {year}
  {2008})}\BibitemShut {NoStop}%
\bibitem [{\citenamefont {Chandrasekhar}(1931)}]{Chandrasekhar1931a}%
  \BibitemOpen
  \bibfield  {author} {\bibinfo {author} {\bibfnamefont {S.}~\bibnamefont
  {Chandrasekhar}},\ }\href {\doibase 10.1080/14786443109461710} {\bibfield
  {journal} {\bibinfo  {journal} {The London, Edinburgh, and Dublin
  Philosophical Magazine and Journal of Science}\ }\textbf {\bibinfo {volume}
  {11}},\ \bibinfo {pages} {592} (\bibinfo {year} {1931})},\ \Eprint
  {http://arxiv.org/abs/https://doi.org/10.1080/14786443109461710}
  {https://doi.org/10.1080/14786443109461710} \BibitemShut {NoStop}%
\bibitem [{\citenamefont {{Chandrasekhar}}(1931)}]{Chandrasekhar1931b}%
  \BibitemOpen
  \bibfield  {author} {\bibinfo {author} {\bibfnamefont {S.}~\bibnamefont
  {{Chandrasekhar}}},\ }\href {\doibase 10.1086/143324} {\bibfield  {journal}
  {\bibinfo  {journal} {\apj}\ }\textbf {\bibinfo {volume} {74}},\ \bibinfo
  {pages} {81} (\bibinfo {year} {1931})}\BibitemShut {NoStop}%
\bibitem [{\citenamefont {Oppenheimer}\ and\ \citenamefont
  {Volkoff}(1939)}]{OppenheimerVolkoff1939}%
  \BibitemOpen
  \bibfield  {author} {\bibinfo {author} {\bibfnamefont {J.~R.}\ \bibnamefont
  {Oppenheimer}}\ and\ \bibinfo {author} {\bibfnamefont {G.~M.}\ \bibnamefont
  {Volkoff}},\ }\href {\doibase 10.1103/PhysRev.55.374} {\bibfield  {journal}
  {\bibinfo  {journal} {Phys. Rev.}\ }\textbf {\bibinfo {volume} {55}},\
  \bibinfo {pages} {374} (\bibinfo {year} {1939})}\BibitemShut {NoStop}%
\bibitem [{\citenamefont {Nilsson}\ and\ \citenamefont
  {Uggla}(2000)}]{NilssonUggla2000}%
  \BibitemOpen
  \bibfield  {author} {\bibinfo {author} {\bibfnamefont {U.~S.}\ \bibnamefont
  {Nilsson}}\ and\ \bibinfo {author} {\bibfnamefont {C.}~\bibnamefont
  {Uggla}},\ }\href {\doibase https://doi.org/10.1006/aphy.2000.6090}
  {\bibfield  {journal} {\bibinfo  {journal} {Annals of Physics}\ }\textbf
  {\bibinfo {volume} {286}},\ \bibinfo {pages} {292} (\bibinfo {year}
  {2000})}\BibitemShut {NoStop}%
\bibitem [{\citenamefont {Buchdahl}(1959)}]{Buchdahl1959}%
  \BibitemOpen
  \bibfield  {author} {\bibinfo {author} {\bibfnamefont {H.~A.}\ \bibnamefont
  {Buchdahl}},\ }\href {\doibase 10.1103/PhysRev.116.1027} {\bibfield
  {journal} {\bibinfo  {journal} {Phys. Rev.}\ }\textbf {\bibinfo {volume}
  {116}},\ \bibinfo {pages} {1027} (\bibinfo {year} {1959})}\BibitemShut
  {NoStop}%
\bibitem [{\citenamefont {Bondi}(1964)}]{Bondi1964}%
  \BibitemOpen
  \bibfield  {author} {\bibinfo {author} {\bibfnamefont {H.}~\bibnamefont
  {Bondi}},\ }\href@noop {} {\bibfield  {journal} {\bibinfo  {journal}
  {Proceedings of the Royal Society of London. Series A. Mathematical and
  Physical Sciences}\ }\textbf {\bibinfo {volume} {282}},\ \bibinfo {pages}
  {303} (\bibinfo {year} {1964})}\BibitemShut {NoStop}%
\bibitem [{\citenamefont {Wald}(1984)}]{Wald1984}%
  \BibitemOpen
  \bibfield  {author} {\bibinfo {author} {\bibfnamefont {R.~M.}\ \bibnamefont
  {Wald}},\ }\href {\doibase 10.7208/chicago/9780226870373.001.0001} {\emph
  {\bibinfo {title} {{General Relativity}}}}\ (\bibinfo  {publisher} {Chicago
  Univ. Pr.},\ \bibinfo {address} {Chicago, USA},\ \bibinfo {year}
  {1984})\BibitemShut {NoStop}%
\bibitem [{\citenamefont {{Schwarzschild}}(1916)}]{Schwarzschild1916b}%
  \BibitemOpen
  \bibfield  {author} {\bibinfo {author} {\bibfnamefont {K.}~\bibnamefont
  {{Schwarzschild}}},\ }in\ \href@noop {} {\emph {\bibinfo {booktitle}
  {Sitzungsberichte der K{\"o}niglich Preussischen Akademie der Wissenschaften
  zu Berlin}}}\ (\bibinfo {year} {1916})\ pp.\ \bibinfo {pages}
  {424--434}\BibitemShut {NoStop}%
\bibitem [{\citenamefont {Andr\'easson}\ and\ \citenamefont
  {Rein}(2007)}]{Andreasson2006}%
  \BibitemOpen
  \bibfield  {author} {\bibinfo {author} {\bibfnamefont {H.}~\bibnamefont
  {Andr\'easson}}\ and\ \bibinfo {author} {\bibfnamefont {G.}~\bibnamefont
  {Rein}},\ }\href {\doibase 10.1088/0264-9381/24/7/008} {\bibfield  {journal}
  {\bibinfo  {journal} {Class. Quant. Grav.}\ }\textbf {\bibinfo {volume}
  {24}},\ \bibinfo {pages} {1809} (\bibinfo {year} {2007})},\ \Eprint
  {http://arxiv.org/abs/gr-qc/0611053} {arXiv:gr-qc/0611053} \BibitemShut
  {NoStop}%
\bibitem [{\citenamefont {Israel}(1966)}]{Israel1966}%
  \BibitemOpen
  \bibfield  {author} {\bibinfo {author} {\bibfnamefont {W.}~\bibnamefont
  {Israel}},\ }\href {\doibase 10.1007/BF02710419} {\bibfield  {journal}
  {\bibinfo  {journal} {Nuovo Cim. B}\ }\textbf {\bibinfo {volume} {44S10}},\
  \bibinfo {pages} {1} (\bibinfo {year} {1966})},\ \bibinfo {note} {[Erratum:
  Nuovo Cim.B 48, 463 (1967)]}\BibitemShut {NoStop}%
\bibitem [{\citenamefont {{Baumgarte}}\ and\ \citenamefont
  {{Rendall}}(1993)}]{Baumgarte1993}%
  \BibitemOpen
  \bibfield  {author} {\bibinfo {author} {\bibfnamefont {T.~W.}\ \bibnamefont
  {{Baumgarte}}}\ and\ \bibinfo {author} {\bibfnamefont {A.~D.}\ \bibnamefont
  {{Rendall}}},\ }\href {\doibase 10.1088/0264-9381/10/2/014} {\bibfield
  {journal} {\bibinfo  {journal} {Classical and Quantum Gravity}\ }\textbf
  {\bibinfo {volume} {10}},\ \bibinfo {pages} {327} (\bibinfo {year}
  {1993})}\BibitemShut {NoStop}%
\bibitem [{\citenamefont {Mars}\ \emph {et~al.}(1996)\citenamefont {Mars},
  \citenamefont {Martin-Prats},\ and\ \citenamefont {Senovilla}}]{Mars1996}%
  \BibitemOpen
  \bibfield  {author} {\bibinfo {author} {\bibfnamefont {M.}~\bibnamefont
  {Mars}}, \bibinfo {author} {\bibfnamefont {M.~M.}\ \bibnamefont
  {Martin-Prats}}, \ and\ \bibinfo {author} {\bibfnamefont {J.~M.~M.}\
  \bibnamefont {Senovilla}},\ }\href {\doibase 10.1016/0375-9601(96)00391-X}
  {\bibfield  {journal} {\bibinfo  {journal} {Phys. Lett. A}\ }\textbf
  {\bibinfo {volume} {218}},\ \bibinfo {pages} {147} (\bibinfo {year}
  {1996})},\ \Eprint {http://arxiv.org/abs/gr-qc/0202003} {arXiv:gr-qc/0202003}
  \BibitemShut {NoStop}%
\bibitem [{\citenamefont {Karageorgis}\ and\ \citenamefont
  {Stalker}(2008)}]{Karageorgis2007}%
  \BibitemOpen
  \bibfield  {author} {\bibinfo {author} {\bibfnamefont {P.}~\bibnamefont
  {Karageorgis}}\ and\ \bibinfo {author} {\bibfnamefont {J.~G.}\ \bibnamefont
  {Stalker}},\ }\href {\doibase 10.1088/0264-9381/25/19/195021} {\bibfield
  {journal} {\bibinfo  {journal} {Class. Quant. Grav.}\ }\textbf {\bibinfo
  {volume} {25}},\ \bibinfo {pages} {195021} (\bibinfo {year} {2008})},\
  \Eprint {http://arxiv.org/abs/0707.3632} {arXiv:0707.3632 [gr-qc]}
  \BibitemShut {NoStop}%
\bibitem [{\citenamefont {{Mazur}}\ and\ \citenamefont
  {{Mottola}}(2023)}]{MazurMottola2023}%
  \BibitemOpen
  \bibfield  {author} {\bibinfo {author} {\bibfnamefont {P.~O.}\ \bibnamefont
  {{Mazur}}}\ and\ \bibinfo {author} {\bibfnamefont {E.}~\bibnamefont
  {{Mottola}}},\ }\href {\doibase 10.3390/universe9020088} {\bibfield
  {journal} {\bibinfo  {journal} {Universe}\ }\textbf {\bibinfo {volume} {9}},\
  \bibinfo {eid} {88} (\bibinfo {year} {2023})}\BibitemShut {NoStop}%
\bibitem [{\citenamefont {Danielsson}\ \emph {et~al.}(2017)\citenamefont
  {Danielsson}, \citenamefont {Dibitetto},\ and\ \citenamefont
  {Giri}}]{Danielssonetal2017}%
  \BibitemOpen
  \bibfield  {author} {\bibinfo {author} {\bibfnamefont {U.~H.}\ \bibnamefont
  {Danielsson}}, \bibinfo {author} {\bibfnamefont {G.}~\bibnamefont
  {Dibitetto}}, \ and\ \bibinfo {author} {\bibfnamefont {S.}~\bibnamefont
  {Giri}},\ }\href {\doibase 10.1007/JHEP10(2017)171} {\bibfield  {journal}
  {\bibinfo  {journal} {JHEP}\ }\textbf {\bibinfo {volume} {10}},\ \bibinfo
  {pages} {171} (\bibinfo {year} {2017})},\ \Eprint
  {http://arxiv.org/abs/1705.10172} {arXiv:1705.10172 [hep-th]} \BibitemShut
  {NoStop}%
\bibitem [{\citenamefont {Arrechea}\ \emph {et~al.}(2021)\citenamefont
  {Arrechea}, \citenamefont {Barcel\'o}, \citenamefont {Carballo-Rubio},\ and\
  \citenamefont {Garay}}]{Arrecheaetal2021}%
  \BibitemOpen
  \bibfield  {author} {\bibinfo {author} {\bibfnamefont {J.}~\bibnamefont
  {Arrechea}}, \bibinfo {author} {\bibfnamefont {C.}~\bibnamefont {Barcel\'o}},
  \bibinfo {author} {\bibfnamefont {R.}~\bibnamefont {Carballo-Rubio}}, \ and\
  \bibinfo {author} {\bibfnamefont {L.~J.}\ \bibnamefont {Garay}},\ }\href
  {\doibase 10.1103/PhysRevD.104.084071} {\bibfield  {journal} {\bibinfo
  {journal} {Phys. Rev. D}\ }\textbf {\bibinfo {volume} {104}},\ \bibinfo
  {pages} {084071} (\bibinfo {year} {2021})},\ \Eprint
  {http://arxiv.org/abs/2105.11261} {arXiv:2105.11261 [gr-qc]} \BibitemShut
  {NoStop}%
\bibitem [{\citenamefont {Arrechea}\ \emph {et~al.}(2024)\citenamefont
  {Arrechea}, \citenamefont {Barcel\'o}, \citenamefont {Carballo-Rubio},\ and\
  \citenamefont {Garay}}]{Arrecheaetal2023}%
  \BibitemOpen
  \bibfield  {author} {\bibinfo {author} {\bibfnamefont {J.}~\bibnamefont
  {Arrechea}}, \bibinfo {author} {\bibfnamefont {C.}~\bibnamefont {Barcel\'o}},
  \bibinfo {author} {\bibfnamefont {R.}~\bibnamefont {Carballo-Rubio}}, \ and\
  \bibinfo {author} {\bibfnamefont {L.~J.}\ \bibnamefont {Garay}},\ }\href
  {\doibase 10.1103/PhysRevD.109.104056} {\bibfield  {journal} {\bibinfo
  {journal} {Phys. Rev. D}\ }\textbf {\bibinfo {volume} {109}},\ \bibinfo
  {pages} {104056} (\bibinfo {year} {2024})},\ \Eprint
  {http://arxiv.org/abs/2310.12668} {arXiv:2310.12668 [gr-qc]} \BibitemShut
  {NoStop}%
\bibitem [{\citenamefont {Martin-Garcia}\ \emph {et~al.}(2021)\citenamefont
  {Martin-Garcia}, \citenamefont {Garc{\'\i}a-Parrado}, \citenamefont
  {Stecchina}, \citenamefont {Wardell}, \citenamefont {Pitrou}, \citenamefont
  {Brizuela} \emph {et~al.}}]{xAct}%
  \BibitemOpen
  \bibfield  {author} {\bibinfo {author} {\bibfnamefont {J.~M.}\ \bibnamefont
  {Martin-Garcia}}, \bibinfo {author} {\bibfnamefont {A.}~\bibnamefont
  {Garc{\'\i}a-Parrado}}, \bibinfo {author} {\bibfnamefont {A.}~\bibnamefont
  {Stecchina}}, \bibinfo {author} {\bibfnamefont {B.}~\bibnamefont {Wardell}},
  \bibinfo {author} {\bibfnamefont {C.}~\bibnamefont {Pitrou}}, \bibinfo
  {author} {\bibfnamefont {D.}~\bibnamefont {Brizuela}},  \emph {et~al.},\
  }\href@noop {} {\bibfield  {journal} {\bibinfo  {journal} {{\tt
  \url{http://www.xact.es}}}\ } (\bibinfo {year} {latest version Oct.
  2021})}\BibitemShut {NoStop}%
\bibitem [{\citenamefont {Misner}\ and\ \citenamefont
  {Sharp}(1964)}]{Misner1964}%
  \BibitemOpen
  \bibfield  {author} {\bibinfo {author} {\bibfnamefont {C.~W.}\ \bibnamefont
  {Misner}}\ and\ \bibinfo {author} {\bibfnamefont {D.~H.}\ \bibnamefont
  {Sharp}},\ }\href {\doibase 10.1103/PhysRev.136.B571} {\bibfield  {journal}
  {\bibinfo  {journal} {Phys. Rev.}\ }\textbf {\bibinfo {volume} {136}},\
  \bibinfo {pages} {B571} (\bibinfo {year} {1964})}\BibitemShut {NoStop}%
\bibitem [{\citenamefont {{Hernandez}}\ and\ \citenamefont
  {{Misner}}(1966)}]{HernandezMisner1966}%
  \BibitemOpen
  \bibfield  {author} {\bibinfo {author} {\bibfnamefont {J.}~\bibnamefont
  {{Hernandez}}, \bibfnamefont {Walter~C.}}\ and\ \bibinfo {author}
  {\bibfnamefont {C.~W.}\ \bibnamefont {{Misner}}},\ }\href {\doibase
  10.1086/148525} {\bibfield  {journal} {\bibinfo  {journal} {\apj}\ }\textbf
  {\bibinfo {volume} {143}},\ \bibinfo {pages} {452} (\bibinfo {year}
  {1966})}\BibitemShut {NoStop}%
\bibitem [{\citenamefont {{Bardeen}}(1968)}]{Bardeen1968}%
  \BibitemOpen
  \bibfield  {author} {\bibinfo {author} {\bibfnamefont {J.}~\bibnamefont
  {{Bardeen}}},\ }in\ \href@noop {} {\emph {\bibinfo {booktitle} {Proceedings
  of the 5th International Conference on Gravitation and the Theory of
  Relativity}}}\ (\bibinfo {year} {1968})\ p.~\bibinfo {pages} {87}\BibitemShut
  {NoStop}%
\bibitem [{\citenamefont {Hayward}(1996)}]{Hayward1994}%
  \BibitemOpen
  \bibfield  {author} {\bibinfo {author} {\bibfnamefont {S.~A.}\ \bibnamefont
  {Hayward}},\ }\href {\doibase 10.1103/PhysRevD.53.1938} {\bibfield  {journal}
  {\bibinfo  {journal} {Phys. Rev.}\ }\textbf {\bibinfo {volume} {D53}},\
  \bibinfo {pages} {1938} (\bibinfo {year} {1996})},\ \Eprint
  {http://arxiv.org/abs/gr-qc/9408002} {arXiv:gr-qc/9408002 [gr-qc]}
  \BibitemShut {NoStop}%
\bibitem [{\citenamefont {Dymnikova}(1992)}]{Dymnikova1992}%
  \BibitemOpen
  \bibfield  {author} {\bibinfo {author} {\bibfnamefont {I.}~\bibnamefont
  {Dymnikova}},\ }\href {\doibase 10.1007/BF00760226} {\bibfield  {journal}
  {\bibinfo  {journal} {Gen. Rel. Grav.}\ }\textbf {\bibinfo {volume} {24}},\
  \bibinfo {pages} {235} (\bibinfo {year} {1992})}\BibitemShut {NoStop}%
\bibitem [{\citenamefont {{Lindblom}}(1984)}]{Lindblom1984}%
  \BibitemOpen
  \bibfield  {author} {\bibinfo {author} {\bibfnamefont {L.}~\bibnamefont
  {{Lindblom}}},\ }\href {\doibase 10.1086/161800} {\bibfield  {journal}
  {\bibinfo  {journal} {\apj}\ }\textbf {\bibinfo {volume} {278}},\ \bibinfo
  {pages} {364} (\bibinfo {year} {1984})}\BibitemShut {NoStop}%
\bibitem [{\citenamefont {Ivanov}(2002)}]{Ivanov2002}%
  \BibitemOpen
  \bibfield  {author} {\bibinfo {author} {\bibfnamefont {B.~V.}\ \bibnamefont
  {Ivanov}},\ }\href {\doibase 10.1103/PhysRevD.65.104011} {\bibfield
  {journal} {\bibinfo  {journal} {Phys. Rev. D}\ }\textbf {\bibinfo {volume}
  {65}},\ \bibinfo {pages} {104011} (\bibinfo {year} {2002})},\ \Eprint
  {http://arxiv.org/abs/gr-qc/0201090} {arXiv:gr-qc/0201090} \BibitemShut
  {NoStop}%
\bibitem [{\citenamefont {Barraco}\ \emph {et~al.}(2003)\citenamefont
  {Barraco}, \citenamefont {Hamity},\ and\ \citenamefont
  {Gleiser}}]{Barraco2003}%
  \BibitemOpen
  \bibfield  {author} {\bibinfo {author} {\bibfnamefont {D.~E.}\ \bibnamefont
  {Barraco}}, \bibinfo {author} {\bibfnamefont {V.~H.}\ \bibnamefont {Hamity}},
  \ and\ \bibinfo {author} {\bibfnamefont {R.~J.}\ \bibnamefont {Gleiser}},\
  }\href {\doibase 10.1103/PhysRevD.67.064003} {\bibfield  {journal} {\bibinfo
  {journal} {Phys. Rev. D}\ }\textbf {\bibinfo {volume} {67}},\ \bibinfo
  {pages} {064003} (\bibinfo {year} {2003})}\BibitemShut {NoStop}%
\bibitem [{\citenamefont {Boehmer}\ and\ \citenamefont
  {Harko}(2006)}]{Boehmer2006}%
  \BibitemOpen
  \bibfield  {author} {\bibinfo {author} {\bibfnamefont {C.~G.}\ \bibnamefont
  {Boehmer}}\ and\ \bibinfo {author} {\bibfnamefont {T.}~\bibnamefont
  {Harko}},\ }\href {\doibase 10.1088/0264-9381/23/22/023} {\bibfield
  {journal} {\bibinfo  {journal} {Class. Quant. Grav.}\ }\textbf {\bibinfo
  {volume} {23}},\ \bibinfo {pages} {6479} (\bibinfo {year} {2006})},\ \Eprint
  {http://arxiv.org/abs/gr-qc/0609061} {arXiv:gr-qc/0609061} \BibitemShut
  {NoStop}%
\bibitem [{\citenamefont {Andr\'easson}(2008)}]{Andreasson2007}%
  \BibitemOpen
  \bibfield  {author} {\bibinfo {author} {\bibfnamefont {H.}~\bibnamefont
  {Andr\'easson}},\ }\href {\doibase 10.1016/j.jde.2008.05.010} {\bibfield
  {journal} {\bibinfo  {journal} {J. Diff. Eq.}\ }\textbf {\bibinfo {volume}
  {245}},\ \bibinfo {pages} {2243} (\bibinfo {year} {2008})},\ \Eprint
  {http://arxiv.org/abs/gr-qc/0702137} {arXiv:gr-qc/0702137} \BibitemShut
  {NoStop}%
\bibitem [{\citenamefont {Urbano}\ and\ \citenamefont
  {Veerm\"ae}(2019)}]{Urbano2018}%
  \BibitemOpen
  \bibfield  {author} {\bibinfo {author} {\bibfnamefont {A.}~\bibnamefont
  {Urbano}}\ and\ \bibinfo {author} {\bibfnamefont {H.}~\bibnamefont
  {Veerm\"ae}},\ }\href {\doibase 10.1088/1475-7516/2019/04/011} {\bibfield
  {journal} {\bibinfo  {journal} {JCAP}\ }\textbf {\bibinfo {volume} {04}},\
  \bibinfo {pages} {011} (\bibinfo {year} {2019})},\ \Eprint
  {http://arxiv.org/abs/1810.07137} {arXiv:1810.07137 [gr-qc]} \BibitemShut
  {NoStop}%
\bibitem [{\citenamefont {Carballo-Rubio}(2018)}]{Carballo-Rubio2017}%
  \BibitemOpen
  \bibfield  {author} {\bibinfo {author} {\bibfnamefont {R.}~\bibnamefont
  {Carballo-Rubio}},\ }\href {\doibase 10.1103/PhysRevLett.120.061102}
  {\bibfield  {journal} {\bibinfo  {journal} {Phys. Rev. Lett.}\ }\textbf
  {\bibinfo {volume} {120}},\ \bibinfo {pages} {061102} (\bibinfo {year}
  {2018})},\ \Eprint {http://arxiv.org/abs/1706.05379} {arXiv:1706.05379
  [gr-qc]} \BibitemShut {NoStop}%
\bibitem [{\citenamefont {Wyman}(1949)}]{Wyman1939}%
  \BibitemOpen
  \bibfield  {author} {\bibinfo {author} {\bibfnamefont {M.}~\bibnamefont
  {Wyman}},\ }\href {\doibase 10.1103/PhysRev.75.1930} {\bibfield  {journal}
  {\bibinfo  {journal} {Phys. Rev.}\ }\textbf {\bibinfo {volume} {75}},\
  \bibinfo {pages} {1930} (\bibinfo {year} {1949})}\BibitemShut {NoStop}%
\bibitem [{\citenamefont {Einstein}(1917)}]{Einstein1917}%
  \BibitemOpen
  \bibfield  {author} {\bibinfo {author} {\bibfnamefont {A.}~\bibnamefont
  {Einstein}},\ }\href {https://adsabs.harvard.edu/pdf/1917SPAW.......142E}
  {\bibfield  {journal} {\bibinfo  {journal} {Sitzungs. K\"onig. Preuss.
  Akad.}\ }\textbf {\bibinfo {volume} {\hspace{-1mm}}},\ \bibinfo {pages} {142}
  (\bibinfo {year} {1917})}\BibitemShut {NoStop}%
\bibitem [{\citenamefont {Jampolski}\ and\ \citenamefont
  {Rezzolla}(2024)}]{Jampolski2023}%
  \BibitemOpen
  \bibfield  {author} {\bibinfo {author} {\bibfnamefont {D.}~\bibnamefont
  {Jampolski}}\ and\ \bibinfo {author} {\bibfnamefont {L.}~\bibnamefont
  {Rezzolla}},\ }\href {\doibase 10.1088/1361-6382/ad2317} {\bibfield
  {journal} {\bibinfo  {journal} {Class. Quant. Grav.}\ }\textbf {\bibinfo
  {volume} {41}},\ \bibinfo {pages} {065014} (\bibinfo {year} {2024})},\
  \Eprint {http://arxiv.org/abs/2310.13946} {arXiv:2310.13946 [gr-qc]}
  \BibitemShut {NoStop}%
\bibitem [{\citenamefont {Cardoso}\ and\ \citenamefont
  {Pani}(2019)}]{CardosoPani2019}%
  \BibitemOpen
  \bibfield  {author} {\bibinfo {author} {\bibfnamefont {V.}~\bibnamefont
  {Cardoso}}\ and\ \bibinfo {author} {\bibfnamefont {P.}~\bibnamefont {Pani}},\
  }\href {\doibase 10.1007/s41114-019-0020-4} {\bibfield  {journal} {\bibinfo
  {journal} {Living Rev. Rel.}\ }\textbf {\bibinfo {volume} {22}},\ \bibinfo
  {pages} {4} (\bibinfo {year} {2019})},\ \Eprint
  {http://arxiv.org/abs/1904.05363} {arXiv:1904.05363 [gr-qc]} \BibitemShut
  {NoStop}%
\bibitem [{\citenamefont {Ayzenberg}\ \emph {et~al.}(2023)\citenamefont
  {Ayzenberg} \emph {et~al.}}]{Ayzenbergetal2023}%
  \BibitemOpen
  \bibfield  {author} {\bibinfo {author} {\bibfnamefont {D.}~\bibnamefont
  {Ayzenberg}} \emph {et~al.},\ }\href@noop {} {\  (\bibinfo {year} {2023})},\
  \Eprint {http://arxiv.org/abs/2312.02130} {arXiv:2312.02130 [astro-ph.HE]}
  \BibitemShut {NoStop}%
\bibitem [{\citenamefont {Rosa}\ and\ \citenamefont
  {Pi\c{c}arra}(2020)}]{Rosa2020}%
  \BibitemOpen
  \bibfield  {author} {\bibinfo {author} {\bibfnamefont {J.~a.~L.}\
  \bibnamefont {Rosa}}\ and\ \bibinfo {author} {\bibfnamefont {P.}~\bibnamefont
  {Pi\c{c}arra}},\ }\href {\doibase 10.1103/PhysRevD.102.064009} {\bibfield
  {journal} {\bibinfo  {journal} {Phys. Rev. D}\ }\textbf {\bibinfo {volume}
  {102}},\ \bibinfo {pages} {064009} (\bibinfo {year} {2020})},\ \Eprint
  {http://arxiv.org/abs/2006.09854} {arXiv:2006.09854 [gr-qc]} \BibitemShut
  {NoStop}%
\bibitem [{\citenamefont {Alho}\ \emph
  {et~al.}(2022{\natexlab{a}})\citenamefont {Alho}, \citenamefont {Nat\'ario},
  \citenamefont {Pani},\ and\ \citenamefont {Raposo}}]{Alho2021}%
  \BibitemOpen
  \bibfield  {author} {\bibinfo {author} {\bibfnamefont {A.}~\bibnamefont
  {Alho}}, \bibinfo {author} {\bibfnamefont {J.}~\bibnamefont {Nat\'ario}},
  \bibinfo {author} {\bibfnamefont {P.}~\bibnamefont {Pani}}, \ and\ \bibinfo
  {author} {\bibfnamefont {G.}~\bibnamefont {Raposo}},\ }\href {\doibase
  10.1103/PhysRevD.105.044025} {\bibfield  {journal} {\bibinfo  {journal}
  {Phys. Rev. D}\ }\textbf {\bibinfo {volume} {105}},\ \bibinfo {pages}
  {044025} (\bibinfo {year} {2022}{\natexlab{a}})},\ \bibinfo {note} {[Erratum:
  Phys.Rev.D 105, 129903 (2022)]},\ \Eprint {http://arxiv.org/abs/2107.12272}
  {arXiv:2107.12272 [gr-qc]} \BibitemShut {NoStop}%
\bibitem [{\citenamefont {Alho}\ \emph
  {et~al.}(2022{\natexlab{b}})\citenamefont {Alho}, \citenamefont {Nat\'ario},
  \citenamefont {Pani},\ and\ \citenamefont {Raposo}}]{Alho2022}%
  \BibitemOpen
  \bibfield  {author} {\bibinfo {author} {\bibfnamefont {A.}~\bibnamefont
  {Alho}}, \bibinfo {author} {\bibfnamefont {J.}~\bibnamefont {Nat\'ario}},
  \bibinfo {author} {\bibfnamefont {P.}~\bibnamefont {Pani}}, \ and\ \bibinfo
  {author} {\bibfnamefont {G.}~\bibnamefont {Raposo}},\ }\href {\doibase
  10.1103/PhysRevD.106.L041502} {\bibfield  {journal} {\bibinfo  {journal}
  {Phys. Rev. D}\ }\textbf {\bibinfo {volume} {106}},\ \bibinfo {pages}
  {L041502} (\bibinfo {year} {2022}{\natexlab{b}})},\ \Eprint
  {http://arxiv.org/abs/2202.00043} {arXiv:2202.00043 [gr-qc]} \BibitemShut
  {NoStop}%
\bibitem [{\citenamefont {Alho}\ \emph
  {et~al.}(2024{\natexlab{a}})\citenamefont {Alho}, \citenamefont {Nat\'ario},
  \citenamefont {Pani},\ and\ \citenamefont {Raposo}}]{Alho2023}%
  \BibitemOpen
  \bibfield  {author} {\bibinfo {author} {\bibfnamefont {A.}~\bibnamefont
  {Alho}}, \bibinfo {author} {\bibfnamefont {J.}~\bibnamefont {Nat\'ario}},
  \bibinfo {author} {\bibfnamefont {P.}~\bibnamefont {Pani}}, \ and\ \bibinfo
  {author} {\bibfnamefont {G.}~\bibnamefont {Raposo}},\ }\href {\doibase
  10.1103/PhysRevD.109.064037} {\bibfield  {journal} {\bibinfo  {journal}
  {Phys. Rev. D}\ }\textbf {\bibinfo {volume} {109}},\ \bibinfo {pages}
  {064037} (\bibinfo {year} {2024}{\natexlab{a}})},\ \Eprint
  {http://arxiv.org/abs/2306.16584} {arXiv:2306.16584 [gr-qc]} \BibitemShut
  {NoStop}%
\bibitem [{\citenamefont {Alho}\ \emph
  {et~al.}(2024{\natexlab{b}})\citenamefont {Alho}, \citenamefont {Nat\'ario},
  \citenamefont {Pani},\ and\ \citenamefont {Raposo}}]{Alho2023b}%
  \BibitemOpen
  \bibfield  {author} {\bibinfo {author} {\bibfnamefont {A.}~\bibnamefont
  {Alho}}, \bibinfo {author} {\bibfnamefont {J.}~\bibnamefont {Nat\'ario}},
  \bibinfo {author} {\bibfnamefont {P.}~\bibnamefont {Pani}}, \ and\ \bibinfo
  {author} {\bibfnamefont {G.}~\bibnamefont {Raposo}},\ }\href {\doibase
  10.1088/1361-6382/ad1e4b} {\bibfield  {journal} {\bibinfo  {journal} {Class.
  Quant. Grav.}\ }\textbf {\bibinfo {volume} {41}},\ \bibinfo {pages} {073002}
  (\bibinfo {year} {2024}{\natexlab{b}})},\ \Eprint
  {http://arxiv.org/abs/2307.03146} {arXiv:2307.03146 [gr-qc]} \BibitemShut
  {NoStop}%
\bibitem [{\citenamefont {Barraco}\ and\ \citenamefont
  {Hamity}(2002)}]{Barraco2002}%
  \BibitemOpen
  \bibfield  {author} {\bibinfo {author} {\bibfnamefont {D.}~\bibnamefont
  {Barraco}}\ and\ \bibinfo {author} {\bibfnamefont {V.~H.}\ \bibnamefont
  {Hamity}},\ }\href {\doibase 10.1103/PhysRevD.65.124028} {\bibfield
  {journal} {\bibinfo  {journal} {Phys. Rev. D}\ }\textbf {\bibinfo {volume}
  {65}},\ \bibinfo {pages} {124028} (\bibinfo {year} {2002})}\BibitemShut
  {NoStop}%
\bibitem [{\citenamefont {Hawking}(1970)}]{Hawking1970}%
  \BibitemOpen
  \bibfield  {author} {\bibinfo {author} {\bibfnamefont {S.}~\bibnamefont
  {Hawking}},\ }\href {\doibase 10.1007/BF01649448} {\bibfield  {journal}
  {\bibinfo  {journal} {Commun. Math. Phys.}\ }\textbf {\bibinfo {volume}
  {18}},\ \bibinfo {pages} {301} (\bibinfo {year} {1970})}\BibitemShut
  {NoStop}%
\bibitem [{\citenamefont {Hawking}\ and\ \citenamefont
  {Ellis}(2023)}]{Hawking1973}%
  \BibitemOpen
  \bibfield  {author} {\bibinfo {author} {\bibfnamefont {S.~W.}\ \bibnamefont
  {Hawking}}\ and\ \bibinfo {author} {\bibfnamefont {G.~F.~R.}\ \bibnamefont
  {Ellis}},\ }\href {\doibase 10.1017/9781009253161} {\emph {\bibinfo {title}
  {{The Large Scale Structure of Space-Time}}}},\ Cambridge Monographs on
  Mathematical Physics\ (\bibinfo  {publisher} {Cambridge University Press},\
  \bibinfo {year} {2023})\BibitemShut {NoStop}%
\bibitem [{\citenamefont {Wong}(2011)}]{Wong2010}%
  \BibitemOpen
  \bibfield  {author} {\bibinfo {author} {\bibfnamefont {W.~W.-Y.}\
  \bibnamefont {Wong}},\ }\href {\doibase 10.1088/0264-9381/28/21/215008}
  {\bibfield  {journal} {\bibinfo  {journal} {Class. Quant. Grav.}\ }\textbf
  {\bibinfo {volume} {28}},\ \bibinfo {pages} {215008} (\bibinfo {year}
  {2011})},\ \Eprint {http://arxiv.org/abs/1011.3029} {arXiv:1011.3029
  [math-ph]} \BibitemShut {NoStop}%
\bibitem [{\citenamefont {Babichev}\ \emph {et~al.}(2008)\citenamefont
  {Babichev}, \citenamefont {Mukhanov},\ and\ \citenamefont
  {Vikman}}]{Babichev2007}%
  \BibitemOpen
  \bibfield  {author} {\bibinfo {author} {\bibfnamefont {E.}~\bibnamefont
  {Babichev}}, \bibinfo {author} {\bibfnamefont {V.}~\bibnamefont {Mukhanov}},
  \ and\ \bibinfo {author} {\bibfnamefont {A.}~\bibnamefont {Vikman}},\ }\href
  {\doibase 10.1088/1126-6708/2008/02/101} {\bibfield  {journal} {\bibinfo
  {journal} {JHEP}\ }\textbf {\bibinfo {volume} {02}},\ \bibinfo {pages} {101}
  (\bibinfo {year} {2008})},\ \Eprint {http://arxiv.org/abs/0708.0561}
  {arXiv:0708.0561 [hep-th]} \BibitemShut {NoStop}%
\bibitem [{\citenamefont {Geroch}(2011)}]{Geroch2010}%
  \BibitemOpen
  \bibfield  {author} {\bibinfo {author} {\bibfnamefont {R.}~\bibnamefont
  {Geroch}},\ }\href@noop {} {\bibfield  {journal} {\bibinfo  {journal} {AMS/IP
  Stud. Adv. Math.}\ }\textbf {\bibinfo {volume} {49}},\ \bibinfo {pages} {59}
  (\bibinfo {year} {2011})},\ \Eprint {http://arxiv.org/abs/1005.1614}
  {arXiv:1005.1614 [gr-qc]} \BibitemShut {NoStop}%
\bibitem [{\citenamefont {Barcel\'o}\ \emph {et~al.}(2022)\citenamefont
  {Barcel\'o}, \citenamefont {S\'anchez}, \citenamefont {Garc\'\i{}a-Moreno},\
  and\ \citenamefont {Jannes}}]{Barcelo2022}%
  \BibitemOpen
  \bibfield  {author} {\bibinfo {author} {\bibfnamefont {C.}~\bibnamefont
  {Barcel\'o}}, \bibinfo {author} {\bibfnamefont {J.~E.}\ \bibnamefont
  {S\'anchez}}, \bibinfo {author} {\bibfnamefont {G.}~\bibnamefont
  {Garc\'\i{}a-Moreno}}, \ and\ \bibinfo {author} {\bibfnamefont
  {G.}~\bibnamefont {Jannes}},\ }\href {\doibase
  10.1140/epjc/s10052-022-10275-3} {\bibfield  {journal} {\bibinfo  {journal}
  {Eur. Phys. J. C}\ }\textbf {\bibinfo {volume} {82}},\ \bibinfo {pages} {299}
  (\bibinfo {year} {2022})},\ \Eprint {http://arxiv.org/abs/2201.11072}
  {arXiv:2201.11072 [gr-qc]} \BibitemShut {NoStop}%
\bibitem [{\citenamefont {Curiel}(2017)}]{Curiel2014}%
  \BibitemOpen
  \bibfield  {author} {\bibinfo {author} {\bibfnamefont {E.}~\bibnamefont
  {Curiel}},\ }\href {\doibase 10.1007/978-1-4939-3210-8_3} {\bibfield
  {journal} {\bibinfo  {journal} {Einstein Stud.}\ }\textbf {\bibinfo {volume}
  {13}},\ \bibinfo {pages} {43} (\bibinfo {year} {2017})},\ \Eprint
  {http://arxiv.org/abs/1405.0403} {arXiv:1405.0403 [physics.hist-ph]}
  \BibitemShut {NoStop}%
\end{thebibliography}%

\end{document}